\documentclass[journal]{IEEEtran}
\usepackage{amsmath,amsfonts,amssymb}
\usepackage[scr=boondox]{mathalpha}
\usepackage{algorithmic}
\usepackage{algorithm}
\usepackage{array}
\usepackage{textcomp}
\usepackage{stfloats}
\usepackage{url}
\usepackage{verbatim}

\usepackage{bm}
\usepackage{enumitem}
\usepackage{mathtools}
\usepackage{cite}
\usepackage{hyperref}
\usepackage{orcidlink}
\usepackage{xcolor}
\usepackage{graphicx}

\hypersetup{colorlinks=true, linkcolor=black, citecolor=black, urlcolor=black}
\usepackage{booktabs,tabularx}

\ifCLASSINFOpdf
  \graphicspath{{figs_r1/}}
  \DeclareGraphicsExtensions{.pdf,.jpeg,.png}
\else
  \graphicspath{{figs_r1/}}
  \DeclareGraphicsExtensions{.eps}
\fi

\newcommand{\sinc}{\operatorname{sinc}}

\newtheorem{remark}{Remark}

\begin{document}

\title{A Unified Pulse-Shaped OFDM Framework for Chirp-Domain Waveforms: Continuous-Time Modeling and Practical I/O Analysis}%

\author{
  \vspace{0.3cm}
  Yating Jiang,~\IEEEmembership{Graduate Student Member,~IEEE}, Hai~Lin,~\IEEEmembership{Senior Member,~IEEE}, \\Yi-Han Chiang,~\IEEEmembership{Member,~IEEE}, Jun Tong,~\IEEEmembership{Member,~IEEE}%
  \thanks{\copyright~2026 IEEE. Personal use of this material is permitted. Permission from IEEE must be obtained for all other uses, in any current or future media, including reprinting/republishing this material for advertising or promotional purposes, creating new collective works, for resale or redistribution to servers or lists, or reuse of any copyrighted component of this work in other works.}%
  \thanks{This paper will be presented in part at the IEEE GLOBECOM 2026, Macau, China \cite{jiang_afdm_gc_2026}.}%
  \thanks{Y. Jiang, H. Lin, and Y.-H. Chiang are with the Department of Electrical and Electronic Systems Engineering, Graduate School of Engineering, Osaka Metropolitan University, Sakai, Osaka 599-8531, Japan (e-mail: sw25630b@st.omu.ac.jp, hai.lin@ieee.org, chiang@omu.ac.jp).}%
  \thanks{J. Tong is with the School of Engineering, University of Wollongong, Wollongong,
    NSW 2522, Australia (e-mail: jtong@uow.edu.au).}%
  \vspace{-0.3cm}
}

\markboth{IEEE Journal on Selected Areas in Communications}%
{Jiang \MakeLowercase{\textit{et al.}}: A Unified Pulse-Shaped OFDM Framework for Chirp-Domain Waveforms}

\maketitle

\begin{abstract}
  A unified framework for chirp-domain waveforms, including orthogonal chirp division multiplexing (OCDM) and affine frequency division multiplexing (AFDM), is developed. Their continuous-time representations are shown to fall within the conventional Weyl-Heisenberg (WH) framework for multicarrier waveforms, with the root chirp as the prototype pulse. Since the root chirp has constant envelope and is transparent to subcarrier orthogonality, these waveforms can be further interpreted as pulse-shaped (PS) orthogonal frequency division multiplexing (OFDM) signals, whose power spectral density is derived analytically. The derived spectrum reveals that implementations based on the discrete affine Fourier transform rely on \emph{sub-Nyquist} samples and exhibit frequency aliasing. We prove that the corresponding aliased chirps are only conditionally orthogonal, whereas sample-wise root-Nyquist pulse shaping of the discrete-time AFDM (DT-AFDM) sequence produces mutually orthogonal pulse-shaped chirps, resulting in the pulse-shaped AFDM (PS-AFDM) waveform. We then derive an exact waveform-level input-output (I/O) relation for PS-AFDM over delay-Doppler (DD) channels, showing that the effective channel at a practical receiver is generally \emph{not} a superposition of pure path-wise DD components. 
  Waveform simulations verify the derived relation to machine precision, while the conventional sequence-level I/O relation for DT-AFDM exhibits a substantial mismatch with waveform behavior for practical channels with continuous-valued delays.
  \end{abstract}

\begin{IEEEkeywords} Chirp-domain, Delay-Doppler domain, Weyl-Heisenberg framework, OFDM, PS-OFDM, OCDM, AFDM, PS-AFDM, F-AFDM, ODDM, I/O relation
\end{IEEEkeywords}

\section{Introduction}
\IEEEPARstart{N}{ext-gen\-er\-a\-tion} wireless systems, including sixth-gen\-er\-a\-tion (6G) mobile networks, vehicle-to-every\-thing (V2X) communications, high-speed railway links, and millimeter-wave/tera\-hertz (mmWave/THz) systems, are increasingly expected to operate in scenarios with very high user or scatterer mobility. Such mobility induces a large Doppler spread and shortens the channel coherence time. In conventional orthogonal frequency division multiplexing (OFDM) systems, this often manifests as increased inter-car\-rier interference. Once the Doppler spread normalized by the subcarrier spacing is no longer sufficiently small, subcarriers may effectively lose their frequency-do\-main orthogonality. This behavior is consistent with the fact that OFDM is typically designed under an approximately linear time-in\-vari\-ant (LTI) channel assumption, whereas under high mobility the physical channel is more appropriately modeled as a linear time-vary\-ing (LTV) system. Consequently, it becomes necessary to represent the channel in domains that better capture the LTV structure, and to design waveforms and transceivers accordingly.

Within the classical wide-sense stationary uncorrelated scattering model \cite{bello_ltvcm_tcom_1963,tse_fwc_2005,Hlawatsch_wcortvc_2011}, an LTV channel is statistically characterized by the scattering function, whose marginals yield the power-delay profile and the Doppler power spectrum. However, in highly time-varying wideband regimes, these statistical averages may not explicitly reveal how transmitted symbols are influenced by individual propagation paths. To obtain a more structured view, it is useful to adopt a delay-Doppler (DD) representation, in which the channel is described by a DD spreading function that quantifies how signal energy is shifted by each path's delay and Doppler before reaching the receiver. This representation makes channel effects more transparent, thereby motivating waveform and equalization designs that explicitly account for the channel structure in the DD domain.

Recent waveform designs for LTV channels can be broadly categorized, from the DD perspective, into two representative approaches. The first approach comprises DD-domain waveforms, which aim to exploit structured DD-domain channel behavior and the associated diversity by multiplexing symbols directly on a DD grid, as exemplified by orthogonal time-frequency space (OTFS) modulation \cite{hadani_otfs_wcnc_2017} and orthogonal delay-Doppler multiplexing (ODDM) \cite{lin_oddm_twc_2022}. The second approach consists of chirp-domain waveforms, including orthogonal chirp division multiplexing (OCDM) \cite{ouyang_ocdm_tcom_2016} and its generalization, affine frequency division multiplexing (AFDM) \cite{bemani_afdm_twc_2023}, which adopt a chirp-domain view by employing chirp-based ``subcarriers'' instead of sinusoidal ones in OFDM.
Since chirps exhibit a swept instantaneous frequency, a path-induced delay shift can be translated into a frequency shift and combined with the Doppler effect of the path. Such chirp-domain waveforms aim to improve robustness to the channel's DD effects while maintaining an OFDM-like block structure and comparable implementation complexity.

Since the physical units of delay and Doppler are time and frequency, respectively, the DD domain is naturally a time-frequency (TF) domain. As a result, it has been shown that DD-domain waveforms can be interpreted within the well-known Weyl-Heisenberg (WH) framework for multicarrier (MC) waveforms, with ODDM constituting a pulse-shaped OFDM (PS-OFDM) scheme that extends the existing WH framework through a (bi)orthogonal WH-subset design
\cite{lin_ddop_gc_2022,lin_mcevo_tcom_2026,lin_primer_spawc_2025}. Recent studies of ODDM waveforms over general physical channels can be found in \cite{tong_oddmphy_tcom_2024,tong_oddmhsphy_twc_2026,pan_cpfreeoddm_twc_2026}.

On the other hand, although orthogonal chirps play a central role in chirp-based waveforms, a systematic framework for understanding their fundamental properties remains elusive. Furthermore, while some studies have addressed the continuous-time representation of chirp-domain waveforms \cite{ouyang_ocdm_tcom_2016,bemani_foldedchirp_wcl_2024,yin_afafdm_jsac_2026}, much of the existing literature is based on discrete-time sequences generated by the discrete Fresnel transform (DFnT) \cite{ouyang_ocdm_tcom_2016} or the discrete affine Fourier transform (DAFT) \cite{bemani_afdm_twc_2023,erseghe_aft_tcom_2005}. Given that the transmitted waveform is ultimately a \emph{continuous-time} signal propagating through a physical channel characterized by \emph{continuous-valued} delay and Doppler, this discrepancy can lead to a non-negligible mismatch between sequence-level analyses and actual channel behavior.

In this paper, we develop a unified framework for chirp-domain waveforms, including OCDM and AFDM.
The framework connects the continuous-time WH representation of chirp-domain waveforms to their spectral properties, sampled implementations, and practical input-output (I/O) behavior over DD channels. The contributions summarized below form a \emph{logically connected progression} rather than independent observations, with each step motivating the next and establishing why an accurate waveform-level I/O relation is \emph{necessary} rather than merely more detailed.

\begin{itemize}

  \item Based on their continuous-time representations, we show that chirp-domain waveforms belong to the conventional WH framework for MC waveforms. In this formulation, the root chirp serves as a constant-envelope prototype pulse that is transparent to subcarrier orthogonality. Hence, these waveforms can be interpreted as PS-OFDM signals. Employing affine chirps as the so-called ``subcarriers'' therefore does \emph{not} render these waveforms a fundamentally distinct class of MC waveforms, since their subcarrier orthogonality arises from the same mechanism as in sinusoid-based ones. The corresponding AFDM waveform coincides with the affine Fourier transform multicarrier (AFT-MC) signal in \cite{stojanovic_aftmc_twc_2009}, which we equivalently term continuous-time AFDM (CT-AFDM), and includes the fractional Fourier transform (FrFT)-based multicarrier waveform in \cite{martone_frft_tcom_2001} as a special case.

  \item Within the developed framework, the power spectral density (PSD) of these waveforms is derived analytically. It reveals that the bandwidth of the AFT-MC/CT-AFDM waveform can substantially exceed the sampling rate adopted in DAFT-based implementations, which explains why the discrete-time AFDM (DT-AFDM) sequence produced by the DAFT \cite{bemani_afdm_twc_2023,erseghe_aft_tcom_2005} consists of \emph{sub-Nyquist} samples of the AFT-MC/CT-AFDM waveform and therefore corresponds to a frequency-aliased waveform.
  \item We then revisit the previously proposed ideal AFDM (I-AFDM) waveform \cite{bemani_foldedchirp_wcl_2024}, hitherto regarded as the ideal continuous-time interpretation of the DT-AFDM sequence and constructed using aliased chirps. We prove that these chirps are only conditionally orthogonal, implying that I-AFDM does not provide a generally applicable continuous-time realization of the DT-AFDM sequence. Since the sequence itself cannot be transmitted directly, we instead form its continuous-time waveform by sample-wise root-Nyquist pulse shaping, which yields mutually orthogonal pulse-shaped chirps without the conditional parameter restrictions of I-AFDM. The resulting transmittable waveform is pulse-shaped AFDM (PS-AFDM), or equivalently filtered AFDM (F-AFDM).

  \item For the resulting PS-AFDM waveform, we derive the exact waveform-level I/O relation over DD channels. Unlike the conventional DT-AFDM sequence-level I/O relation, the derived relation accounts for the complete transceiver chain and links the physical propagation paths to the equivalent channel taps through a kernel given by the cross-ambiguity function of the transmit and receive pulses. Our analysis shows that, after transmit pulse shaping and receive filtering, the effective channel is generally not a superposition of pure path-wise DD components, as commonly assumed in the literature. That assumption amounts to replacing the correct kernel by a \emph{delta} function, which is generally unjustified.

  \item Simulations verify the derived PSD and numerically corroborate the exact waveform-level I/O relation across several pulse, channel, and mobility settings, with the predicted received signal agreeing with waveform simulations to double-precision roundoff. We then investigate the validity regime of the conventional DT-AFDM sequence-level I/O relation by characterizing its modeling mismatch with respect to fractional delay. We also compare the waveform-generation complexity with that of ODDM, another PS-OFDM scheme, for the same number of transmitted symbols and subcarrier spacing.

\end{itemize}

Note that this paper proposes no pilot pattern, channel estimator, equalizer, or detector, and claims no bit error rate advantage over existing waveforms, since such designs must first rest on an accurate I/O relation that describes the transceiver they are built for.

The remainder of this paper is organized as follows. Section~\ref{sec:mc_principles} introduces the WH framework for MC waveforms, and Section~\ref{sec:psofdm_framework} develops the unified PS-OFDM framework for chirp-domain waveforms and derives their PSD. Section~\ref{sec:waveforms} investigates sub-Nyquist implementations and the orthogonality of the corresponding aliased chirps, and Section~\ref{sec:io_relation} derives the I/O relation of the PS-AFDM waveform over DD channels. Section~\ref{sec:simulation} reports the numerical study and Section~\ref{sec:conclusion} concludes the paper.

\textbf{Notation:} Uppercase and lowercase boldface letters represent matrices and column vectors, respectively.
The superscripts $\mathsf{T}$, $\mathsf{H}$, and $*$ indicate transpose, conjugate transpose, and complex conjugate.
The operator $\star$ denotes convolution. The normalized sinc function is defined as
$\sinc(\zeta)\triangleq \frac{\sin(\pi\zeta)}{\pi\zeta}$. The function $\Pi_T(t)$ is a unit-amplitude rectangular pulse with support $[0,T]$, and
$\delta_{i,j}$ denotes the Kronecker delta, while $\delta(\cdot)$ denotes the Dirac delta.
$\mathcal A_{g,\tilde g}(\tau,\nu)$ denotes the cross-ambiguity function between $g(t)$ and $\tilde g(t)$, given by
\begin{align*}
  \mathcal A_{g,\tilde g}(\tau,\nu)
   & = \langle g(t), \tilde g(t-\tau)e^{j2\pi \nu (t-\tau)} \rangle                  \\
   & = \int_{-\infty}^{+\infty} g(t)\,\tilde g^*(t-\tau)e^{-j2\pi \nu (t-\tau)}\,dt.
\end{align*}

\section{MC Waveform Principles}
\label{sec:mc_principles}

Given the symbol index $i \in \mathbb{Z}$, a \emph{real-valued} passband digital modulation waveform with carrier frequency $f_c$ and phase $\theta_c$ is typically represented as \cite{lathi_mdacs_1998,proakis_dc4th_2000,madhow_fdc_2008}:
{\small
\begin{align*}
  x_{pb}(t) = \sqrt{2} \sum_i  A_i \cos\!\left(2\pi \int_0^t \left(f_c + f_i(\tilde t)\right) d\tilde t + \theta_c + \theta_i\right) \rho_i(t),
\end{align*}}%
where the amplitude $A_i$, the phase $\theta_i$, the instantaneous frequency offset $f_i(\tilde t)$, and the finite-duration pulse $\rho_i(t)$ can be used for modulation. The corresponding \emph{complex-valued} baseband waveform is $x(t) = \sum_i x_i(t)$, where the $i$-th symbol is defined as:
\begin{equation}\label{eq:x_it}
  x_i(t) = A_i e^{j\theta_i} e^{j2\pi \int_0^t f_i(\tilde t) d\tilde t} \rho_i(t) \triangleq X[i] g_i(t).
\end{equation}
This consists of two components: an information-bearing \emph{digital data symbol} (a complex number) $X[i] = A_i e^{j\theta_i}$, drawn from a signaling alphabet, and a finite-energy, continuous-time function $g_i(t) = e^{j2\pi \int_0^t f_i(\tilde t) d\tilde t} \rho_i(t)$, referred to as the \emph{transmit (modulating) pulse} or filter.

In the receiver, a correlator performing cross-correlation with a \emph{receive pulse} $\gamma_i(t)$ is typically used to extract the $i$-th digital data symbol as:
\begin{align}\label{eq:rx_correlator}
  Y[i] = \int_{-\infty}^{+\infty} y(t) \gamma_i^*(t) dt,
\end{align}
where $y(t)$ is the received waveform. The correlator can be implemented by a matched filter \cite{lathi_mdacs_1998,proakis_dc4th_2000,madhow_fdc_2008}, whose output sample at an appropriate time yields $Y[i]$.

To avoid inter-symbol interference, the transmit pulses are required to be mutually orthogonal or paired with a biorthogonal set of receive pulses. Hence, a digital modulation waveform can be defined by a set of \emph{(bi)orthogonal pulses}, also referred to as \emph{basis functions}.

For MC systems, the symbol index $i$ is replaced by a time index $m$ and a frequency index $n$. Assuming that $N$ is even, the MC waveform is given by \cite{lin_mcevo_tcom_2026,lin_primer_spawc_2025,matz_tff_spm_2013,sahin_mct_cst_2014}:
\begin{align}\label{eq:xt}
  x(t) = \sum_{m=0}^{M-1} \sum_{n=-\frac{N}{2}}^{\frac{N}{2}-1} X[m,n] \overbrace{e^{j2\pi n \Delta F (t - m \Delta T)} g(t - m \Delta T)}^{g_{m,n}(t)},
\end{align}
where $N$ and $M$ are the numbers of subcarriers and MC symbols, respectively. $\Delta T$ and $\Delta F$ are the time and frequency resolutions, and $g(t)$ is called the \emph{prototype pulse}. In the literature, $\Delta T$ is also known as the symbol interval, while $\Delta F$ is referred to as the fundamental frequency or subcarrier spacing. Since $e^{j2\pi n \Delta F t}$ is often called a \emph{subcarrier} or \emph{tone}, the transmit pulses $g_{m,n}(t)$ in \eqref{eq:xt} are truncated or pulse-shaped (windowed) subcarriers, corresponding to a \emph{rectangular} or \emph{non-rectangular} $g(t)$, respectively.

For such an MC waveform characterized by a TF grid $(\Delta T, \Delta F)$ and a prototype pulse $g(t)$, a design framework based on the WH frame theory is well known \cite{matz_tff_spm_2013,sahin_mct_cst_2014}, as $g_{m,n}(t)$ in \eqref{eq:xt} is the WH or Gabor function adopted in TF signal analysis \cite{grochenig_ftfa_2001}.

A WH set $\{\Delta T, \Delta F, g(t)\}$ is said to be \emph{orthogonal} given the inner product:
\begin{align}
  \langle g_{m,n}(t), g_{m', n'}(t)\rangle = E_g\delta_{m,m'}\delta_{n,n'},
\end{align}
where $E_g=\langle g,g\rangle$ is the energy of the transmit pulse.
Let the corresponding receive pulse be another WH function $\gamma_{m,n}(t) = \gamma(t-m\Delta T)e^{j2\pi n\Delta F (t-m\Delta T)}$. A pair of WH sets $\{\Delta T, \Delta F, g(t)\}$ and $\{\Delta T, \Delta F, \gamma(t)\}$ is said to be \emph{biorthogonal} when:
\begin{align}
  \langle g_{m,n}(t), \gamma_{m', n'}(t)\rangle = E_{g,\gamma}\delta_{m,m'}\delta_{n,n'},
\end{align}
where $E_{g,\gamma}=\langle g,\gamma\rangle$ is the cross-correlation between the transmit and receive pulses.
According to the WH frame theory, (bi)orthogonal WH sets only exist when the joint TF resolution $\Delta R = \Delta T \Delta F \geq 1$. A typical example of such a (bi)orthogonal WH set is that adopted in OFDM, with $\Delta T = T + T_{cp}$, $\Delta F = 1/T$, $g(t) = \Pi_{T+T_{cp}}(t)$, and $\gamma(t) = \Pi_{T}(t)$, where $T_{cp}$ is the length of the cyclic prefix (CP). MC waveforms employing (bi)orthogonal WH functions with non-rectangular $g(t)$ are therefore called PS-OFDM \cite{zhao_psofdm_ewcom_2017}.

\section{PS-OFDM Framework for OCDM/AFDM}
\label{sec:psofdm_framework}

\subsection{OCDM Waveform}

An OCDM signal having $N$ orthogonal chirps can be written as \cite{ouyang_ocdm_tcom_2016}
\begin{align}
  x_{\rm OC}(t)=\sum_{n=0}^{N-1} X[n]\,\psi_n(t), \quad 0\le t<T, \label{eq:ocdm}
\end{align}
where $X[n]$ is the information-bearing digital data symbol,
\begin{align}
  \psi_n(t)=\Pi_T(t)e^{j\frac{\pi}{4}}e^{-j\alpha \pi \left ( t-\frac{n}{\alpha T}\right )^2},
  \label{eq:nth-chirp}
\end{align}
is the $n$-th chirp and $\alpha=\frac{N}{T^2}$ is the chirp rate. The corresponding root chirp is given by
\begin{align}
  \psi_0(t)=\Pi_T(t)e^{j\frac{\pi}{4}}e^{-j\alpha \pi t^2}.
\end{align}
It is known that these chirps are orthogonal to each other, evidenced by
\begin{align}
  \int_{0}^{T}\psi_n(t)\,\psi^*_{n'}(t)\,dt=T\delta_{n,n'}.
  \label{eq:chirp-orthogonality}
\end{align}

The implementation of the OCDM signal is based on the DFnT \cite{ouyang_ocdm_tcom_2016}, as the discrete OCDM signal is given by sampling the waveform in \eqref{eq:ocdm} at a rate of $\frac{N}{T}$ as
\begin{align}
  x_{\rm OC}[k] & =  x_{\rm OC}\left(k\frac{T}{N}\right) \nonumber                          \\
                 & =  \sum_{n=0}^{N-1} X[n] e^{\frac{j\pi}{4} } e^{-j\frac{\pi}{N} ( k-n)^2},
\end{align}
for $0 \le k \le N-1$, where $e^{\frac{j\pi}{4} } e^{-j\frac{\pi}{N} (k-n)^2}$ is exactly the $(k,n)$ entry of the $N\times N$ inverse DFnT (IDFnT) matrix up to a scaling factor.

\subsection{AFDM Waveform}
The foundation of the AFDM signal is the affine Fourier transform (AFT), namely the linear canonical transform \cite{healy_lct_2016}, given by
\begin{align}
  X(u)=\int_{-\infty}^{+\infty} x(t)\mathcal K(t,u)dt, \label{eq:aft}
\end{align}
where the kernel
\begin{align}
  \mathcal K(t,u)= \frac{1}{\sqrt{2\pi|b|}}
  e^{-j\left(\frac{a}{2b}u^2+\frac{1}{b}ut+\frac{d}{2b}t^2\right)},\label{eq:aft-kernel}
\end{align}
has four parameters $(a, b, c, d)$ that satisfy the constraint $ad-bc=1$. Without loss of generality, we assume $b>0$ hereafter.
It is known that $x(t)$ can be recovered as
\begin{align}
  x(t)=\int_{-\infty}^{+\infty} X(u)\mathcal K^*(t,u)du, \label{eq:iaft}
\end{align}
using the inverse AFT, which is also an AFT with parameters $(d, -b, -c, a)$.

By sampling \eqref{eq:aft} in both the time domain $t$ and the transform domain $u$, with intervals $\Delta t$ and $\Delta u$, respectively, the discrete AFT (DAFT) and the corresponding inverse DAFT (IDAFT) can be defined under the condition \cite{pei_faft_tsp_2000}
\begin{align}\label{eq:dtdurela}
  \Delta t \Delta u =\frac{2\pi b}{N},
\end{align}
where $N$ denotes the number of samples over $u$. For the commonly adopted symmetric sampling grid with $\Delta t=\Delta u$, the transform parameter $b$ follows
\begin{align}
  b = \frac{N \Delta u^2}{2\pi}.
\end{align}

In \cite{bemani_afdm_twc_2023}, the IDAFT is used to generate the DT-AFDM sequence as
\begin{align}\label{eq:afdm-seq}
  x_{\rm AF}[k] & =\frac{1}{\sqrt{N}}
  \sum_{n=0}^{N-1}X[n]e^{j2\pi \left(c_2 n^2+\frac{nk}{N}+ c_1 k^2\right)},
\end{align}
for $0 \le k \le N-1$, $X[n]\triangleq X(n\Delta u)$ and
\begin{align}
  c_1 & =  \frac{d}{4\pi b} \Delta t^2, \\
  c_2 & = \frac{a}{4\pi b} \Delta u^2.
\end{align}

By comparing \eqref{eq:afdm-seq} with \eqref{eq:iaft}, one can see that the corresponding CT-AFDM signal can be obtained by considering sampling over the transform domain $u$ only.
Substituting $u=n\Delta u$ into \eqref{eq:iaft} for $0\le t<T$ and replacing the integral by a Riemann sum yields
\begin{align}
  \label{eq:afdm-continuous-approx}
  x_{\rm AF}(t)
  \approx \sum_{n=0}^{N-1} X[n]
  \underbrace{\mathcal K^*(t,n\Delta u)\Delta u }_{\triangleq \dot \phi_n(t)},
\end{align}
where
\begin{align}
  \dot \phi_n(t)
   & =\frac{\Delta u}{\sqrt{2\pi|b|}} \Pi_T(t) e^{j\left(\frac{a}{2b}(n\Delta u)^2+\frac{1}{b}(n\Delta u)t+\frac{d}{2b}t^2\right)}, \nonumber            \\
   & = \frac{1}{\sqrt{N}} \Pi_T(t) e^{j2\pi \left(c_2n^2+\frac{n}{N\Delta t} t+c_1\left(\frac{t}{\Delta t}\right)^2\right)}. \label{eq:afdm-ideal-basis}
\end{align}

Bearing in mind that $\Delta t=\frac{T}{N}$, comparison of
\eqref{eq:afdm-ideal-basis} with \eqref{eq:nth-chirp} shows that OCDM
is a special case of AFDM with $c_1=-\frac{1}{2N}$, up to a
subcarrier-dependent phase associated with $c_2$ that can be absorbed
into $X[n]$.

Note that the CT-AFDM in \eqref{eq:afdm-continuous-approx} coincides, up to the parameterization of the chirp rate, with the AFT-MC waveform with a rectangular pulse in \cite{stojanovic_aftmc_twc_2009}, and includes the FrFT-based multicarrier waveform in \cite{martone_frft_tcom_2001} as a special case. By contrast, the scheme in \cite{erseghe_aft_tcom_2005}, referred to as DAFT-OFDM in \cite{bemani_afdm_twc_2023}, is inherently discrete-time and shares with DT-AFDM the DAFT-based sequence generation and the chirp-periodic prefix (CPP), differing in the design criterion for the chirp parameters.
Hereafter, we adopt this AFT-MC/CT-AFDM signal
\begin{align}
\label{eq:cd-continuous-time}
x_{\rm AF}(t)
= \sum_{n=0}^{N-1} X[n] \dot{\phi}_n(t),
\end{align}
as the general form of chirp-domain waveforms in the subsequent discussion and use the terms AFT-MC and CT-AFDM interchangeably.

In this chirp-based MC formulation, the use of affine chirps as the so-called ``subcarriers,'' rather than the sinusoidal subcarriers employed in the conventional MC waveform in \eqref{eq:xt}, has often led chirp-based MC waveforms to be regarded as a fundamentally distinct class of MC waveforms. However, as will be demonstrated in the following, this apparent distinction does not imply a different underlying waveform structure.

\subsection{WH and PS-OFDM Frameworks}
By absorbing the term $\frac{1}{\sqrt{N}}e^{j2\pi c_2 n^2}$ into $X[n]$ as $\dot X[n]=\frac{1}{\sqrt{N}}e^{j2\pi c_2 n^2}X[n]$, \eqref{eq:cd-continuous-time} can be rewritten as
\begin{align}
  \label{eq:cd-continuous-time-1}
  x_{\rm AF}(t)
  = \sum_{n=0}^{N-1} \dot X[n] \phi_n(t),
\end{align}
where
\begin{align}
  \phi_n(t) & =  \Pi_T(t)e^{j2\pi c_1 N^2\left(\frac{t}{T}\right)^2}  e^{j2\pi \frac{n}{T} t}. \label{eq:cd-ideal-basis}
\end{align}
Comparing \eqref{eq:cd-continuous-time-1} with \eqref{eq:xt}, one can see that
\eqref{eq:cd-continuous-time-1} corresponds exactly to a \emph{conventional sinusoidal-based} MC symbol, and the basis functions in \eqref{eq:cd-ideal-basis} take the form of WH functions, where $\Delta T = T$, $\Delta F =1/T$, and the prototype pulse becomes
\begin{align}\label{eq:pp_afdm}
  g_{\rm AF}(t) & =\Pi_T(t)e^{j2\pi c_1 N^2\left(\frac{t}{T}\right)^2},
\end{align}
which is precisely the root chirp.

One can observe that this chirp-based prototype pulse is a constant-envelope signal, which does \emph{not} affect the orthogonality among the basis functions, because
\begin{align}
  \label{eq:afdm-ct-orth}
  \int_{0}^{T}\phi_n(t)\phi^*_{n'}(t)\,dt & =\int_{0}^{T}g_{\rm AF}(t)g^*_{\rm AF}(t)e^{j2\pi \frac{n-n'}{T} t} dt, \nonumber \\
                                          & =\int_{0}^{T}e^{j2\pi \frac{n-n'}{T} t}dt, \nonumber                              \\ &=T\delta_{n,n'}.
\end{align}
In other words, $g_{\rm AF}(t)$ can form an orthogonal WH set over the TF grid $\left(T, \frac{1}{T}\right)$.

Recall that when $g(t)$ is non-rectangular and forms an orthogonal WH set over the TF grid $\left(\Delta T, \Delta F\right)$, the MC waveform in \eqref{eq:xt} becomes a PS-OFDM signal.
Consequently, chirp-domain waveforms not only fall within the WH framework but can also be interpreted as PS-OFDM signals with a constant-envelope prototype pulse.

\subsection{PSD Analysis}
Within the above unified PS-OFDM framework, the PSD of chirp-domain waveforms can be derived analytically.

\subsubsection{PSD of PS-OFDM Signals}
For the PSD analysis, we consider the steady-state extension of \eqref{eq:xt}, obtained by letting $m\in\mathbb Z$.
Assume that the information-bearing symbols $X[m,n]$ are i.i.d. with zero mean and average power $\sigma_X^2 = \mathbb E[|X[m,n]|^2]$. It can be proved that $x(t)$ is wide-sense cyclostationary with period $\Delta T$. The time-averaged autocorrelation function of a cyclostationary signal is defined as \cite{proakis_dc4th_2000},
\begin{align}
  \label{eq:bar_Rx_tau}
  \bar R_x(\tau) & =\frac{1}{\Delta T}
  \int_0^{\Delta T}R_x(t,\tau)dt, \nonumber                                             \\
                 & =\frac{1}{\Delta T} \int_0^{\Delta T} \mathbb E[x(t)x^*(t-\tau)] dt.
\end{align}
Using \eqref{eq:xt}, we have
\begin{align}
  \mathbb E[x(t)x^*(t-\tau)] = & \sigma_X^2 \sum_{m} g(t - m \Delta T) g^*(t - \tau - m \Delta T) \nonumber \\ & \times \sum_{n} e^{j2\pi n \Delta F \tau}.  \label{eq:Rx_t_tau}
\end{align}
Substituting \eqref{eq:Rx_t_tau} into \eqref{eq:bar_Rx_tau}, we obtain
\begin{align}
  \bar R_x(\tau)
   & = \frac{\sigma_X^2}{\Delta T} R_g(\tau) \sum_{n=-\frac{N}{2}}^{\frac{N}{2}-1} e^{j2\pi n\Delta F\tau},
\end{align}
where $R_g(\tau)$ is the autocorrelation of the prototype pulse given by
\begin{align}
  R_g(\tau)= \int_{-\infty}^{+\infty} g(t)\,g^*(t-\tau)\,dt.
\end{align}
Applying the Wiener-Khinchin theorem, we have the PSD
\begin{align}
  S_x(f) & = \mathcal{F}\{ \bar{R}_x(\tau) \}, \nonumber                                                                    \\
         & =\frac{\sigma_X^2}{\Delta T} \sum_{n=-\frac{N}{2}}^{\frac{N}{2}-1} |G(f - n \Delta F)|^2, \label{eq:psd-unified}
\end{align}
where $G(f)$ is the Fourier transform of $g(t)$.
It follows from \eqref{eq:psd-unified} that the PSD is determined by the prototype pulse spectrum $|G(f)|^2$, the frequency resolution $\Delta F$, and the time resolution $\Delta T$.

\subsubsection{PSD of AFT-MC/CT-AFDM Signals}
For the AFT-MC/CT-AFDM signal in \eqref{eq:cd-continuous-time-1}, after a centered re-indexing of the subcarrier index to match the unified formulation in \eqref{eq:xt}, and then substituting $g(t)=g_{\rm AF}(t)$, $\Delta T=T$, $\Delta F=1/T$ into \eqref{eq:psd-unified}, we have
\begin{align}
  \label{eq:afdm-psd}
  S_{\rm AF}(f)
  =\frac{\sigma_X^2}{NT} \sum_{n=-\frac{N}{2}}^{\frac{N}{2}-1}
  \left|G_{\rm AF} \left(f-\frac{n+N/2}{T}\right)\right|^2,
\end{align}
where the phase factor $e^{j2\pi c_2 n^2}$ absorbed into $\dot X[n]$ does not affect its second-order statistics, and $G_{\rm AF}(f)$ is
the corresponding prototype pulse spectrum given by
\begin{align}
  \label{eq:root-chirp-spectrum}
  G_{\rm AF}(f)=\int_{0}^{T} e^{j2\pi c_1 N^2\left(\frac{t}{T}\right)^2} e^{-j2\pi ft}\,dt.
\end{align}

Let the span of $G_{\rm AF}(f)$ be $B_{g_{\rm AF}}$, which can be considered as the bandwidth of $g_{\rm AF}(t)$\footnote{Since $g_{\rm AF}(t)$ is time-limited, its spectrum is not strictly bandlimited. The bandwidth can be reasonably defined by ignoring the negligibly small spectral tails beyond the main spectral region.}. 
By evaluating the Fourier integral in \eqref{eq:root-chirp-spectrum} in terms of Fresnel integrals, we obtain $B_{g_{\rm AF}}\approx 2|c_1|N^2/T$, which characterizes the main spectral region $[0,B_{g_{\rm AF}}]$ of the finite-duration chirp (see Appendix~\ref{sec:derivation_of_bandwidth} for a derivation). 
Without loss of generality, we can frequency-shift the root chirp to center its instantaneous-frequency range at zero, so that this region becomes $[-B_{g_{\rm AF}}/2,B_{g_{\rm AF}}/2]$. 
Then, from \eqref{eq:afdm-psd}, the $N$ spectral replicas are spaced by $1/T$, so the approximate bandwidth of the AFT-MC/CT-AFDM signal is given by
\begin{align}
  B_{\rm AF}= \frac{N-1}{T}+B_{g_{\rm AF}}\approx \frac{2|c_1|N^2+N-1}{T}.
\end{align}

\section{I-AFDM and PS-AFDM Waveforms}
\label{sec:waveforms}
Recall that the DT-AFDM sequence has sampling interval $\Delta t=\frac{T}{N}$, or equivalently a sampling frequency of $\frac{N}{T}$. Since $B_{\rm AF}>\frac{N}{T}$ whenever $|c_1|>\frac{1}{2N^2}$, as satisfied by conventional AFDM parameter choices, the DT-AFDM sequence generated by the DAFT consists of sub-Nyquist samples of the AFT-MC/CT-AFDM waveform, which results in frequency aliasing.

\subsection{I-AFDM Waveform}%
\label{sec:iafdm}
Let $C=2N|c_1|$. The I-AFDM waveform, which is regarded as the
ideal continuous-time interpretation of the DT-AFDM sequence, is composed of aliased chirps \cite{bemani_foldedchirp_wcl_2024,yin_afafdm_jsac_2026}
\begin{align}
  \hat \phi_n(t)
  =e^{j2\pi\left(c_2 n^2 + \frac{c_1}{\Delta t^2}t^2 + \frac{n}{T}t - \frac{q}{\Delta t}t\right)},
\end{align}
for $t_{n,q} \le t < t_{n,q+1}$, with
\begin{align}
  t_{n,q}=
  \begin{cases}
    0,                                  & q = 0,            \\[4pt]
    \frac{T}{C}q - \frac{n}{C}\Delta t, & q = 1,2,\ldots,C, \\[6pt]
    T,                                  & q = C+1.
  \end{cases}
\end{align}
However, these aliased chirps are conditionally orthogonal (see Appendix~\ref{sec:orthogonality} for a proof). Consequently, interference may arise in the I-AFDM waveform, making it difficult to employ in communication systems.

\subsection{PS-AFDM (F-AFDM) Waveform}
\label{sec:psafdm}
On the other hand, let $a(t)$ denote a unit-energy pulse-shaping filter. A transmittable waveform is instead obtained by applying sample-wise pulse shaping based on $a(t)$ to the DT-AFDM sequence in \eqref{eq:afdm-seq}, or, equivalently, by passing the sequence in \eqref{eq:afdm-seq} through a filter characterized by $a(t)$.

\begin{remark} A discrete-time sequence is not itself a physical waveform and cannot be transmitted directly. Before carrier modulation and over-the-air transmission, it must be converted into a continuous-time waveform by pulse shaping. It should be noted that generating a discrete-time sequence followed by pulse shaping is only one possible implementation method. The modulation waveform can also be generated directly without explicitly forming the intermediate sequence.
\end{remark}

\begin{remark}
  In MC waveforms, pulse shaping is more complex because each symbol spans multiple time-domain samples usually generated by the inverse fast Fourier transform (IFFT), unlike single-carrier modulation where each sample corresponds to one symbol. It typically involves two steps: \emph{sample-wise} pulse shaping via interpolation filtering (equivalent to that used in the single-carrier case), and subsequent \emph{symbol-wise} pulse shaping via multiplication with a prototype pulse. When the prototype pulse has a constant envelope (e.g., rectangular), sample-wise pulse shaping alone may suffice with appropriate parameter settings \cite{lin_mcevo_tcom_2026,lin_primer_spawc_2025}. Such waveforms are also referred to as filtered OFDM (F-OFDM) \cite{Abdoli_fofdm_spawc_2015}.
\end{remark}

After the sample-wise pulse shaping, we have the PS-AFDM waveform
\begin{align}\label{eq:aa_afdm}
  x_{\rm AF}^{(a)}(t)=\sum_k x_{\rm AF}[k] a\left(t-k\frac{T}{N}\right),
\end{align}
and the corresponding pulse-shaped chirps are given by
\begin{align}
  \label{eq:general-ps}
  \phi_n^{(a)}(t)
   & =\sum_{k=0}^{N-1}\phi_n[k]\,a\left(t-k\frac{T}{N}\right),
\end{align}
where the $n$-th chirp sequence is
\begin{align}
  \label{eq:afdm-sequence}
  \phi_n[k]
   & =\phi_n\!\left(k\frac{T}{N}\right)
  = e^{j2\pi c_1k^2}e^{j2\pi \frac{nk}{N}},
\end{align}
for $0\le k\le N-1$.

Because sample-wise pulse shaping is equivalent to filtering the sequence, the PS-AFDM waveform may also be referred to as an F-AFDM waveform, as in \cite{shen_fafdm_arxiv_2026}. It should be noted that these terms have different meanings in the context of AFDM than in that of OFDM, where PS-OFDM and F-OFDM denote two conceptually distinct waveform constructions \cite{lin_mcevo_tcom_2026}, as discussed above.
For notational convenience, we hereafter use the terms PS-AFDM and F-AFDM interchangeably. This usage differs from that in \cite{yin_psafdm_gc_2024}, where PS-AFDM refers to a windowed I-AFDM variant.

Note that the chirp sequence can be represented as a vector
$
  \bm{\phi}_n=[\phi_n[0], \cdots, \phi_n[N-1]]^{\mathsf T},
$
corresponding to the $n$-th column of the IDAFT matrix.
Since $\langle \bm{\phi}_n, \bm{\phi}_{n'} \rangle=N \delta_{n,n'}$, it is well known that
\begin{align*}
  \langle \phi_n^{(a)}(t), \phi_{n'}^{(a)}(t) \rangle = N \delta_{n,n'},
\end{align*}
when $a(t)$ is a root-Nyquist pulse with Nyquist interval $\frac{T}{N}$ \cite{madhow_fdc_2008}.
Thus, a root-Nyquist pulse-shaping filter with Nyquist interval $\frac{T}{N}$ provides a sufficient construction for preserving mutual orthogonality among these pulse-shaped chirps.
Accordingly, when interpreted as samples of an ideal bandlimited waveform, the DT-AFDM sequence corresponds to a sinc pulse-shaped waveform according to the Nyquist sampling theorem.

\section{I/O Relation over DD Channels}%
\label{sec:io_relation}
Assume the LTV channel consists of $P$ propagation paths. For the $p$-th path, let $\check h_p$, $\tau_p^{\rm phy}$, and $\nu_p$ be its complex gain, physical propagation delay, and Doppler shift, respectively. Let $t_0$ denote the receiver timing reference and define $\tau_p\triangleq\tau_p^{\rm phy}-t_0$ as the path delay relative to the receiver sampling origin.  Also, for notational convenience, the constant delay-dependent Doppler phase is absorbed into the complex path gain by defining $h_p\triangleq\check h_p e^{-j2\pi\nu_p\tau_p}$. This channel model is considered valid during the channel's ``stationary'' interval \cite{bello_ltvcm_tcom_1963, Hlawatsch_wcortvc_2011, lin_mcevo_tcom_2026}. Its representation in the DD domain, also known as the DD spreading function, is given by
\begin{align}\label{eq:sf}
  \tilde h(\tau,\nu)=\sum_{p=1}^{P} h_p \delta (\tau-\tau_p)\delta (\nu-\nu_p),
\end{align}
where $\tau$ and $\nu$ are the delay and Doppler domain variables, respectively. Without loss of generality, we choose $t_0\leq\tau_1^{\rm phy}$ and index the paths such that
\begin{align}
  0\le \tau_1 \le \tau_2 \le \cdots \le \tau_P.
  \label{eq:tau_order}
\end{align}
We assume that the delay spread $\tau_{\rm ds}=\tau_P-\tau_1$ satisfies $\tau_{\rm ds}<T$.
For a transmitted baseband signal $x(t)$, the received baseband signal $y(t)$ is usually expressed as
\begin{align}
  y(t) = \sum_{p=1}^{P} h_p e^{j2\pi \nu_p t} x(t-\tau_p) + z(t), \label{eq:rt_channel}
\end{align}
where $z(t)$ represents additive white Gaussian noise (AWGN) with a PSD of $N_0$.

\subsection{DT-AFDM Sequence-Level I/O Relation}
The conventional DT-AFDM sequence-level I/O relation was derived in \cite{bemani_afdm_twc_2023}, where each path is assumed to have an \emph{on-grid} delay that can be normalized to an integer as $l_p=\tau_p\frac{N}{T}$. Then, the received samples are given as
\begin{align}\label{eq:rk}
  y[k'] & =\sum_l h_{k',l} \, x_{\rm AF}[k'-l]+z[k'],
\end{align}
where $z[k']$ denotes samples of AWGN and
\begin{align}\label{eq:tv_tap}
  h_{k',l} & =\sum_{p=1}^{P} h_p e^{j2\pi \nu _p k'\frac{T}{N}}\delta_{l,l_p},
\end{align}
is the time-varying impulse response of the channel at time $k'$ and delay $l$.
$x_{\rm AF}[\cdot]$ in \eqref{eq:rk} usually includes a CPP with length $L_{\rm CPP}> \lceil \tau_P\frac{N}{T} \rceil -1 $. Under earliest-path alignment, $t_0=\tau_1^{\rm phy}$ and hence $\tau_P=\tau_{\rm ds}$.

In the receiver, after discarding the CPP and letting $\mathbf{y}=[y[0],\cdots, y[N-1]]^{\mathsf T}$, $\mathbf{x}=[x_{\rm AF}[0],\cdots, x_{\rm AF}[N-1]]^{\mathsf T}$, \eqref{eq:rk} can be expressed in matrix form as
\begin{align}\label{eq:r}
  \mathbf{y}
   & =\mathbf{H}\mathbf{x}+\mathbf{z},
\end{align}
where $\mathbf{z}=[z[0],\cdots, z[N-1]]^{\mathsf T}\sim \mathcal{CN}(0,N_0\mathbf{I})$, and
\begin{align}
  \mathbf{H}=\sum_{p=1}^{P}h_p\mathbf{\Gamma}_p\mathbf{\Delta}_{\nu_p}\mathbf{\Pi}^{l_p}.
\end{align}
In this expression, $\mathbf{\Pi}$ denotes the $N\times N$ forward cyclic-shift matrix,
$\mathbf{\Delta}_{\nu_p}\triangleq \mathrm{diag}\left(1, e^{j2\pi \bar\nu_p},\cdots, e^{j2\pi(N-1)\bar\nu_p}\right)$ with $\bar\nu_p\triangleq\nu_p\frac{T}{N}$ the Doppler normalized to the sampling rate $\frac{N}{T}$, and $\mathbf{\Gamma}_p$ is an $N\times N$ diagonal matrix given by
\begin{align}
  \mathbf{\Gamma}_p
   & =\mathrm{diag}\!\left(
  \begin{cases}
    e^{-j2\pi c_1\left(N^2-2N(l_p-n)\right)}, & n<l_p    \\
    1,                                        & n\ge l_p
  \end{cases}
  \right).
\end{align}
By applying the $N \times N$ DAFT matrix $\mathbf A$ to $\mathbf y$, the DAFT-domain I/O relation becomes
\begin{align}\label{eq:yf}
  \mathbf{y}_u
   & =\mathbf{A}\mathbf{y}
  =\underbrace{\mathbf A \mathbf H \mathbf A^{\mathsf H}}_{\mathbf{H}_u}\mathbf{x}_u+\mathbf{A}\mathbf{z},
\end{align}
where $\mathbf{x}_u=[X[0],\cdots, X[N-1]]^{\mathsf T}$.

The DAFT-domain I/O matrix $\mathbf H_u$ plays a fundamental role in AFDM waveform design and performance evaluation, such as in selecting the parameters $c_1$ and $c_2$ to achieve optimal diversity order and reduced pilot overhead \cite{bemani_afdm_twc_2023}.

\subsection{PS-AFDM Waveform-Level I/O Relation}
Although the I/O relation in \eqref{eq:yf} appears appealing, it is derived from the time-domain relation in \eqref{eq:rk}, which, however, does \emph{not} match the behavior of a practical receiver processing chain, as will be explained below.

In practice, with time measured relative to $t_0$, the received real-valued passband signal is
\begin{equation}
  y_{pb}(t) = y_{pb}^S(t) + y_{pb}^I(t) + z_{pb}(t), \label{eq:yt_pb_channel}
\end{equation}
where the desired signal component is
\begin{align}
  y_{pb}^S(t)
  & = \sqrt{2}\,\Re\!\left\{
  \sum_{p=1}^{P}\tilde h_p x(t-\tau_p)
  e^{j2\pi(f_c+\nu_p)(t-\tau_p)}
  \right\} \nonumber \\
  & = \Re\!\left\{y^S(t)\sqrt{2}e^{j2\pi f_ct}\right\},
\end{align}
and
$y^S(t)=\sum_{p=1}^{P} h_p e^{j2\pi \nu_p t} x(t-\tau_p)$ is its complex baseband equivalent with respect to the system carrier reference $\sqrt{2} e^{j2\pi f_c t}$. Here, $\tilde{h}_p$ is a real-valued path gain, $\check h_p\triangleq\tilde{h}_p e^{-j2\pi f_c\tau_p}$ includes the carrier phase associated with the delay, and $h_p=\check h_p e^{-j2\pi\nu_p\tau_p}=\tilde{h}_p e^{-j2\pi(f_c+\nu_p)\tau_p}$ additionally absorbs the constant delay-dependent Doppler phase, consistent with the notation adopted above. $y_{pb}^I(t)$ denotes the out-of-band interference component, and $z_{pb}(t)$ is passband AWGN with a two-sided PSD of $\frac{N_0}{2}$.

To obtain the baseband signal, $y_{pb}(t)$ is first filtered by a real-valued bandpass filter (BPF) $r_{pb}(t)$ to remove $y_{pb}^I(t)$ and is then down-converted \cite{razavi_rfme_2011}. Let $r(t)$ denote the baseband equivalent of $r_{pb}(t)$. After mixing, we obtain
\begin{align}
  \tilde y(t)
  = & y^S(t)\star r(t) + (y^S(t)\star r(t))^*e^{-j4\pi f_c t} \nonumber \\
    & + z(t)\star r(t) + (z(t)\star r(t))^*e^{-j4\pi f_c t},
\end{align}
where $z(t)$ is the baseband equivalent of $z_{pb}(t)$.
Further applying a low-pass filter (LPF) $\tilde r(t)$ to remove the components around $-2f_c$ yields
\begin{align}
  y(t)
   & = y^S(t)\star r(t)\star \tilde r(t) + z(t)\star r(t)\star \tilde r(t), \label{eq:real_bb_model}
\end{align}
where the joint effect of the BPF and the LPF, given by $r(t)\star \tilde r(t)$, can be interpreted as an equivalent baseband receive filter \cite{lin_mcevo_tcom_2026,lin_ddmc_talk_2025}. A block diagram of the standard processing chain can be found in \cite{lin_mcevo_tcom_2026}.

In classical communication theory \cite{lathi_mdacs_1998, proakis_dc4th_2000, madhow_fdc_2008}, the baseband noise $z(t)$ is often modeled as having infinite bandwidth and, consequently, infinite power for mathematical convenience. To justify this model, a subsequent filtering stage is typically introduced, which aligns exactly with \eqref{eq:real_bb_model}. Comparing \eqref{eq:rt_channel} with \eqref{eq:real_bb_model}, it is clear that \eqref{eq:rt_channel} represents an idealized model that neglects the practical receive filtering mentioned above.

Substituting \eqref{eq:aa_afdm} into \eqref{eq:rt_channel}, it can be seen that the time-varying channel tap $h_{k',l}$ in \eqref{eq:rk} is a superposition of pure path-wise DD components, obtained by sampling $y(t)$ with interval $\frac{T}{N}$ for a unit-amplitude Nyquist pulse $a(t)$. Therefore, the effect of \emph{inevitable receive filtering} is not captured in \eqref{eq:rk}, and the statistical characterization of $z[k']$ may not be well justified.

Recall that the pulse-shaped chirps in \eqref{eq:afdm-sequence} are orthogonal to each other when $a(t)$ is a root-Nyquist pulse-shaping filter. Maintaining this orthogonality at the output of the receive filter under an ideal channel naturally suggests employing the same root-Nyquist pulse-shaping filter as the receive filter, which also serves as the matched filter and the \emph{anti-aliasing} filter.

Let us apply the matched filter $a^*(-t)$ to $y(t)$, and let $\mathcal A_{a,a}(\cdot,\cdot)$ denote the ambiguity function of $a(t)$.
The sampled output of the matched filter is given by
\begin{align}
  y^{\rm MF}[k']\label{eq:y_mf_k}
   & =
  \sum_l h_{k',l}^{\rm MF}x_{\rm AF}[k'-l]+z^{\rm MF}[k'],
\end{align}
where the equivalent channel tap is (see Appendix~\ref{sec:derivation_of_channel_tap} for a detailed derivation)
\begin{align}
  \hspace{-3mm}
  h_{k',l}^{\rm MF}
  =
  \sum_{p=1}^{P} h_p\,
  e^{j2\pi \nu_p k'\frac{T}{N}}\,
  \mathcal A_{a,a}(l\frac{T}{N}-\tau_p,-\nu_p).
  \label{eq:tv_tap_mf}
\end{align}
For other transmit and receive pulse pairs, including mismatched pairs, the equivalent channel taps can be obtained by replacing $\mathcal A_{a,a}(\cdot,\cdot)$ in \eqref{eq:tv_tap_mf} with the corresponding cross-ambiguity function. The derivation requires only that the transmitted waveform be generated by sample-wise pulse shaping of a discrete-time sequence.

The ambiguity function in \eqref{eq:tv_tap_mf} thus acts as the kernel linking the physical propagation paths to the equivalent discrete-time channel taps. The physical paths have continuous-valued delays and Doppler shifts, whereas the delay dimension of the equivalent taps is indexed by the integer $l$. The value $\mathcal A_{a,a}(l\frac{T}{N}-\tau_p,-\nu_p)$ specifies the contribution of each path to each tap. When $\nu_p=0$ for all $p$, \eqref{eq:tv_tap_mf} reduces to the classical expression for the equivalent channel taps in \cite{tse_fwc_2005}. In that LTI case, the kernel depends only on delay and is given by the cross-correlation of the transmit and receive filters, which reduces to a sinc function for ideal bandlimited signaling.

Comparing \eqref{eq:tv_tap_mf} with \eqref{eq:tv_tap}, the conventional DT-AFDM sequence-level I/O relation is seen to replace this kernel by a delta function and assume on-grid delays, so that every physical path contributes to exactly one equivalent tap. Since the kernel is generally not a delta function, the effective channel generally \emph{cannot} be expressed as a superposition of pure path-wise DD components, and the I/O relation in \eqref{eq:yf} does not generally describe a practical transceiver.

How far \eqref{eq:yf} departs from the practical I/O relation depends on the transmit and receive pulses, the fractional displacement of the physical delays from the sampling grid, and the Doppler shifts. The departure can persist even for an on-grid path with $\tau_p=l_p\frac{T}{N}$, where $l_p$ is an integer, because the off-center ambiguity-function samples $\mathcal A_{a,a}((l-l_p)\frac{T}{N},-\nu_p)$, $l\neq l_p$, are generally nonzero when $\nu_p\neq0$. Their effects are examined numerically in Section~\ref{sec:io_verification}. 

\begin{remark}\label{rmk:lti}
  For LTI channels with $\nu_p=0$ for all $p$, the equivalent taps incorporate the combined effect of transmit pulse shaping, the physical propagation paths, and receive filtering. The on-grid taps may therefore be interpreted as \emph{correlated virtual paths with on-grid delays} rather than as independent physical propagation paths, whose delays are generally not aligned with the sampling grid. Under this interpretation, the channel model in \eqref{eq:tv_tap} remains valid without explicitly representing the pulse pair. As long as the effective channel remains LTI, analyses based on the on-grid LTI channel model continue to hold after its coefficients are estimated using pilot signals. This is why sequence-based analyses of OFDM and OCDM have historically served well in low-mobility regimes. This reinterpretation does not extend to $\nu_p\neq0$. Matching \eqref{eq:tv_tap_mf} to \eqref{eq:tv_tap} would require the gain $h_p\mathcal A_{a,a}(l\frac{T}{N}-\tau_p,-\nu_p)$, which varies with the tap index $l$, whereas each gain in \eqref{eq:tv_tap} is a single constant. Because this quantity also depends on the transmit and receive pulses, it is a parameter of the effective channel rather than of the physical propagation path. Moreover, the gains obtained in this way for the successive taps of one path are not independent, being samples of a single ambiguity profile fixed by $(\tau_p,\nu_p)$ and the pulse pair. Once the channel becomes time varying, estimating on-grid coefficients therefore no longer justifies \eqref{eq:tv_tap}, since those coefficients cannot be assigned independent delays and Doppler shifts.
\end{remark}

Similarly, after discarding the CPP and letting $\mathbf{y}^{\rm MF}=[y^{\rm MF}[0],\cdots, y^{\rm MF}[N-1]]^{\mathsf T}$, \eqref{eq:y_mf_k} can be written in matrix form as
\begin{align}
  \mathbf{y}^{\rm MF}
   & =\mathbf{H}^{\rm MF}\mathbf{x}+\mathbf{z}^{\rm MF},
\end{align}
where $\mathbf{z}^{\rm MF}=[z^{\rm MF}[0],\cdots, z^{\rm MF}[N-1]]^{\mathsf T}\sim \mathcal{CN}(0,N_0\mathbf{I})$ because $a(t)$ is a root-Nyquist pulse for the Nyquist interval $\frac{T}{N}$ and has unit energy $\int |a(t)|^2 dt =1$.
For transmit and receive filters implemented with finite spans, the ambiguity kernel has finite support in its delay argument. By choosing an appropriate receiver timing reference $t_0$, the nonzero equivalent taps can, without loss of generality, be indexed as $l=0,\cdots,L-1$, where $L$ covers the combined channel and filter span. We assume that $L\le N$ and $L_{\rm CPP}\ge L-1$, so that the equivalent tap support fits within one AFDM block and is covered by the CPP. Substituting $x_{\rm AF}[k]=x_{\rm AF}[N+k]e^{-j2\pi c_1(N^2+2Nk)}$ gives $\acute h_{k',l}^{\rm MF}\triangleq h_{k',l}^{\rm MF}e^{-j2\pi c_1\left(N^2+2N(k'-l)\right)}$, so that $\mathbf H^{\rm MF}\in\mathbb C^{N\times N}$ is formed by CPP folding as
\begin{align*}
  [\mathbf H^{\rm MF}]_{k',k'-l}   & =h_{k',l}^{\rm MF},\quad l=0,\cdots,k',                 \\
  [\mathbf H^{\rm MF}]_{k',N+k'-l} & =\acute h_{k',l}^{\rm MF},\quad l=k'+1,\cdots,L-1.
\end{align*}
Further applying the DAFT matrix $\mathbf A$ to $\mathbf y^{\rm MF}$, we have
\begin{align}
  \mathbf{y}_u^{\rm MF}
   & =\mathbf{A}\mathbf{y}^{\rm MF}
  =\underbrace{\mathbf A \mathbf H^{\rm MF} \mathbf A^{\mathsf H}}_{\mathbf{H}_u^{\rm MF}}\mathbf{x}_u+\mathbf{A}\mathbf{z}^{\rm MF},\label{eq:daft_io}
\end{align}
whose $(n',n)$-th entry follows by substituting the DAFT matrix entries, as
\begin{align}\label{eq:HuMF_entry}
  [\mathbf H_u^{\rm MF}]_{n',n}
   & =\frac{e^{j2\pi c_2(n^2-{n'}^2)}}{N}
  \sum_{k'=0}^{N-1}\sum_{l=0}^{L-1}
  h_{k',l}^{\rm MF} \nonumber \\
   & \quad \times
  e^{j2\pi c_1\left((k'-l)^2-{k'}^2\right)}
  e^{j2\pi\frac{(k'-l)n-k'n'}{N}}.
\end{align}
It is noteworthy that the combination of matched filtering, sampling, and DAFT is equivalent to $N$ correlators based on $\phi_n^{(a)}(t)$ in \eqref{eq:general-ps}, which coincides with \eqref{eq:rx_correlator}.

Overall, \eqref{eq:daft_io} provides the explicit practical I/O relation in the DAFT domain. Specifically, the effective coupling coefficients between $\mathbf{x}_u$ and $\mathbf{y}_u^{\rm MF}$, represented by $\mathbf{H}_u^{\rm MF}$, account for the entire signal chain. Since \eqref{eq:yf} does not describe that chain, \eqref{eq:daft_io} is the relation on which pilot design and channel equalization for practical PS-AFDM systems must be based.

\section{Simulation Results}
\label{sec:simulation}
This section numerically verifies the derived PSD and the exact I/O relation over DD channels, investigates the validity regime of the conventional relation, and compares the waveform-generation complexity of PS-AFDM and ODDM.

In the simulations, the AFDM parameters are set to $T=266.667~\mu$s, $N=1024$,
and the corresponding subcarrier spacing is $\frac{1}{T}=3.75$~kHz. The transmitted symbols are drawn from a $4$-QAM constellation with average power $\sigma_X^2=1$. To generate the PS-AFDM waveform, the DT-AFDM sequence is pulse-shaped by a transmit filter with span $Q\frac{T}{N}$, so that $Q$ is the filter span measured in sampling intervals. For square-root raised-cosine (SRRC) pulse shaping, $\beta$ denotes the roll-off factor.
The multipath channel follows the standard Extended Vehicular A (EVA) model \cite{eva_channel_model}, consisting of $P=9$ paths with delays that are generally off-grid relative to $\frac{T}{N}$. Two diagnostic channels are additionally considered. Quasi-EVA rounds every EVA path delay to the nearest multiple of $\frac{T}{N}$, thereby isolating the Doppler-dependent mismatch under on-grid delays. The single-path channel places one path at $(l_0+\delta)\frac{T}{N}$, thereby isolating the fractional-delay mismatch without multipath accumulation. Unless otherwise stated, $l_0=10$ and $\delta=-0.1$, corresponding to a path delay of $9.9T/N$. For EVA and quasi-EVA, $t_0=0$, which is the earliest path of the tabulated profile. For the diagnostic single-path channel, $t_0=0$ is retained as the fixed receiver sampling origin rather than aligned with its sole path, so that $\tau_1=(l_0+\delta)T/N$ and the fractional-delay effect can be isolated. The carrier frequency is $f_c = 5$~GHz, and the Doppler frequency of the $p$-th path is generated using Jakes' formula $\nu_p = \nu_{\textrm{max}} \cos(\theta_p)$, where the maximum Doppler frequency $\nu_{\textrm{max}}$ is determined by the terminal speed, and $\theta_p$ is uniformly distributed over $[-\pi,\pi]$.

\begin{figure}
  \centering
  \includegraphics[width=8.5cm]{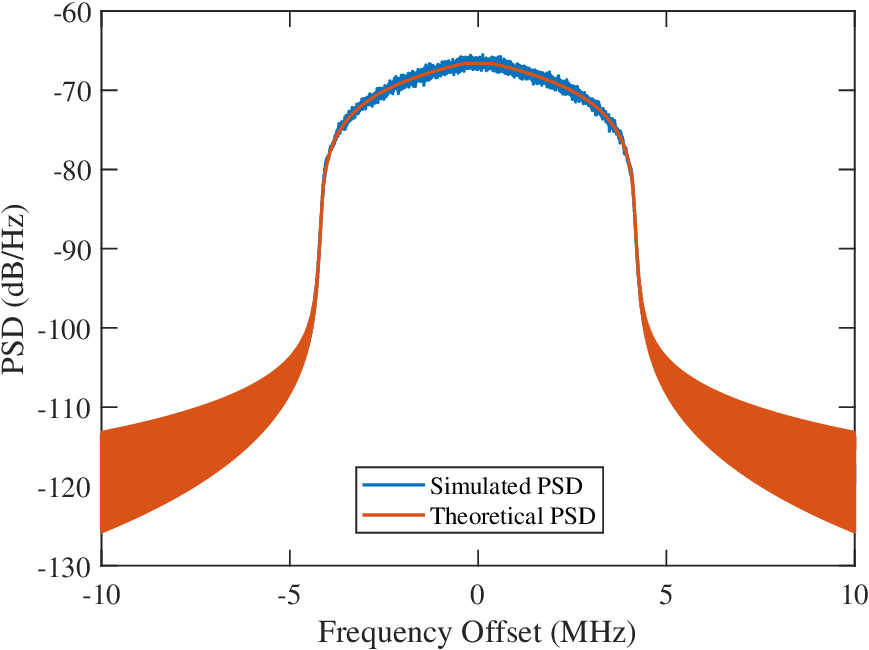}
  \caption{PSD of the CT-AFDM waveform, $c_1=\frac{3}{5N}$, $c_2=\frac{1}{3N}$, $N=1024$.}
  \label{fig:psd}
\end{figure}

\subsection{Verification of the PSD}
We first verify the derived PSD of the CT-AFDM waveform under the developed PS-OFDM framework. Fig.~\ref{fig:psd} compares the simulated PSD with the theoretical PSD derived in Section~\ref{sec:psofdm_framework}, with $c_1=\frac{3}{5N}$ and $c_2=\frac{1}{3N}$.
As shown in the figure, the simulated results closely match the analytical PSD. In particular, the CT-AFDM waveform has an approximate bandwidth of $\frac{2|c_1|N^2+N-1}{T}$, which evaluates to $8.44$~MHz, confirming the correctness of the theoretical characterization of the spectral behavior of the CT-AFDM waveform.

Note that this parameter setting corresponds to $C=2N|c_1|=1.2$, which is not an integer and therefore falls under Case 3 of Appendix~\ref{sec:orthogonality}. The aliased chirps in the I-AFDM waveform are thus not orthogonal for this setting. By contrast, as shown in Section~\ref{sec:psafdm}, the orthogonality of the pulse-shaped chirps in the PS-AFDM waveform follows from the orthogonality of the IDAFT columns together with the root-Nyquist property of $a(t)$, and holds for any $c_1$.

\subsection{Verification of the I/O Relation}
\label{sec:io_verification}
Next, we verify the derived I/O relation for the PS-AFDM waveform over DD channels using the same chirp parameters $c_1=\frac{3}{5N}$ and $c_2=\frac{1}{3N}$.
The accuracy is measured by the normalized mean-square error (NMSE) between the received DAFT-domain signal $\mathbf y^{\rm MF}_u$ predicted by the I/O relation and that obtained from a waveform-level simulation incorporating pulse shaping, the DD channel, matched filtering, sampling, and the DAFT. It is defined as
\begin{align}\label{eq:nmse}
  \mathrm{NMSE}
  =\frac{\mathbb E\!\left[\left\|\widehat{\mathbf y}^{\rm MF}_{u}-\mathbf y^{\rm MF}_{u}\right\|^2\right]}
  {\mathbb E\!\left[\left\|\mathbf y^{\rm MF}_{u}\right\|^2\right]},
\end{align}
where $\widehat{\mathbf y}^{\rm MF}_{u}$ and $\mathbf y^{\rm MF}_{u}$ denote the prediction and waveform-simulation reference, respectively. The expectations are evaluated over the Monte Carlo realizations.

\subsubsection{Validation of the Derived Relation}
\begin{figure}
  \centering
  \includegraphics[width=8.5cm]{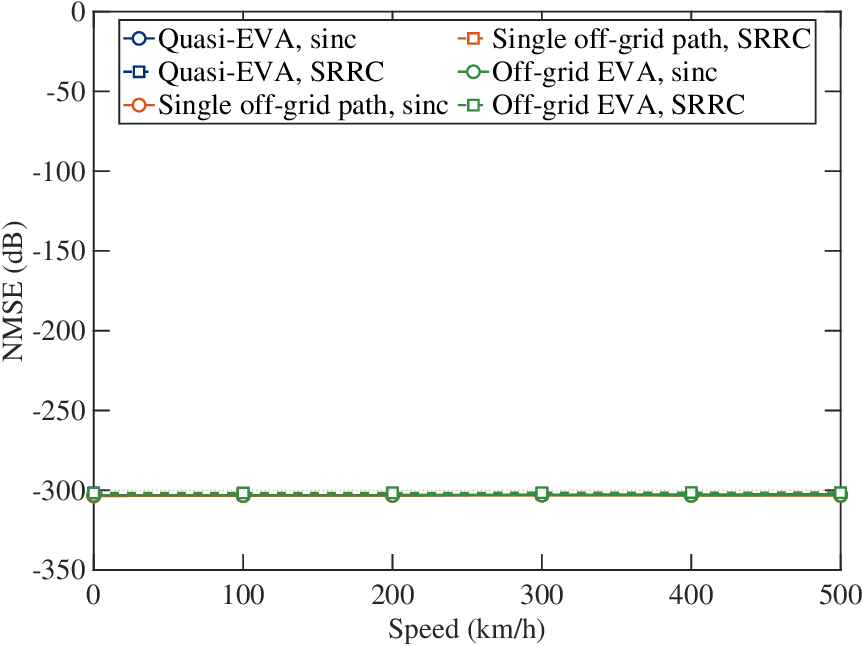}
  \caption{Numerical validation of the PS-AFDM I/O relation versus speed for sinc and SRRC pulse pairs over the standard off-grid EVA channel and the two diagnostic channels. All six curves coincide at the numerical floor.}
  \label{fig:derived_validation}
\end{figure}

Fig.~\ref{fig:derived_validation} validates the derived PS-AFDM waveform-level I/O relation over the two diagnostic channels, quasi-EVA and the single-path channel, as well as the standard off-grid EVA channel, for speeds from 0 to 500~km/h. Two matched pulse pairs are considered, a sinc pair with $Q=128$ and an SRRC pair with $\beta=0.2$ and $Q=12$. Within each pair, the transmit and receive pulses are truncated to the same span, and the ambiguity kernel is evaluated from exactly the same truncated pulse samples used in the waveform-level transceiver simulation. Each of the six pulse-channel combinations remains at approximately $-300$~dB throughout the speed sweep, confirming that \eqref{eq:daft_io} reproduces the simulated transceiver chain to double-precision roundoff. Note that because the waveform simulation evaluates the continuous-time chain on a finite-resolution numerical time grid, this result verifies the derived relation against that numerical waveform simulation rather than quantifying the discretization error relative to the exact continuous-time waveform.

\begin{figure}[!t]
  \centering
  \includegraphics[width=8.5cm]{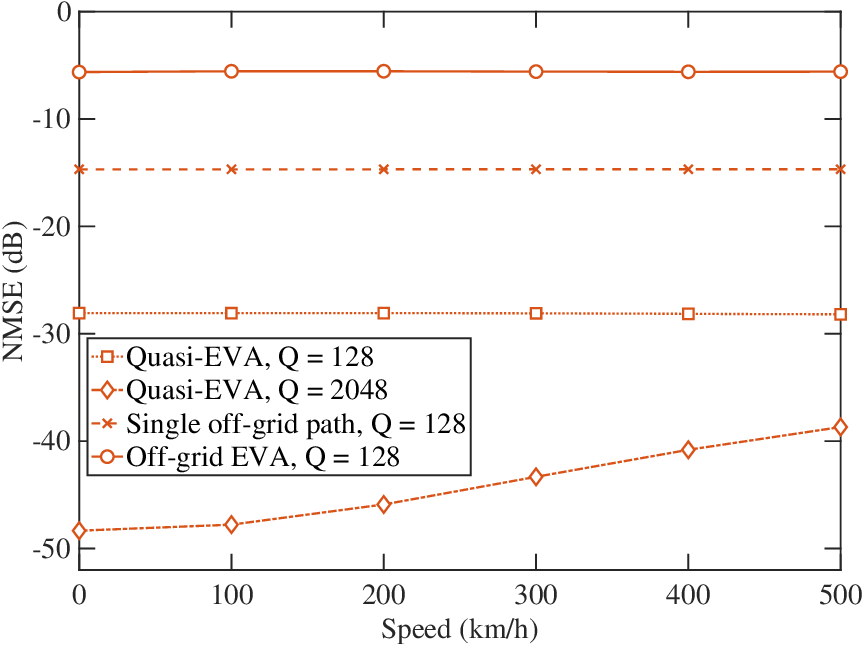}
  \caption{NMSE of the DT-AFDM I/O relation versus speed for the standard off-grid EVA channel and the two diagnostic channels.}
  \label{fig:conventional_eva}
\end{figure}

\subsubsection{Validity Regime of the Conventional Relation}
\begin{figure}
  \centering
  \includegraphics[width=8.5cm]{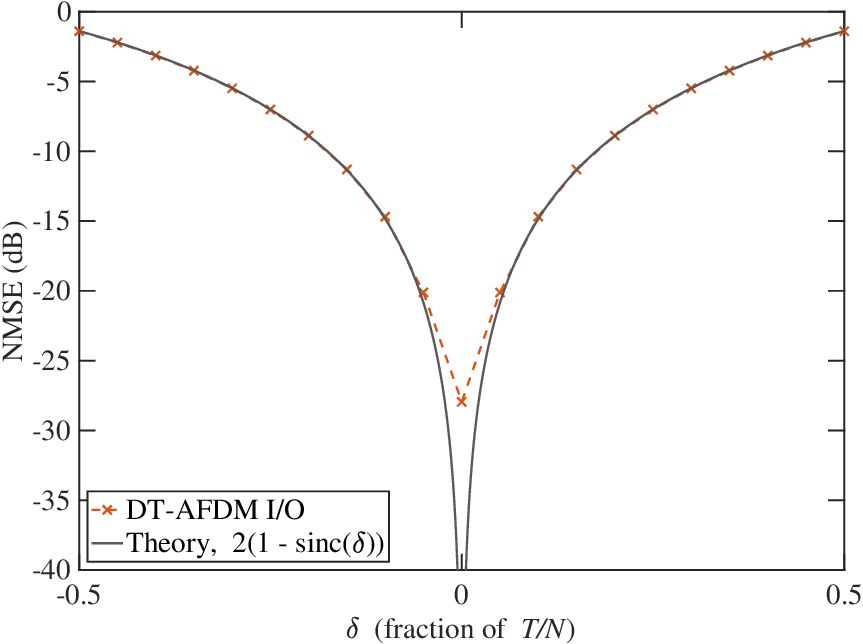}
  \caption{NMSE of the DT-AFDM I/O relation versus fractional delay for the diagnostic single-path channel, together with the theoretical result in \eqref{eq:frac_delay_nmse}.}
  \label{fig:conventional_frac_delay}
\end{figure}

\begin{figure}
  \centering
  \includegraphics[width=8.5cm]{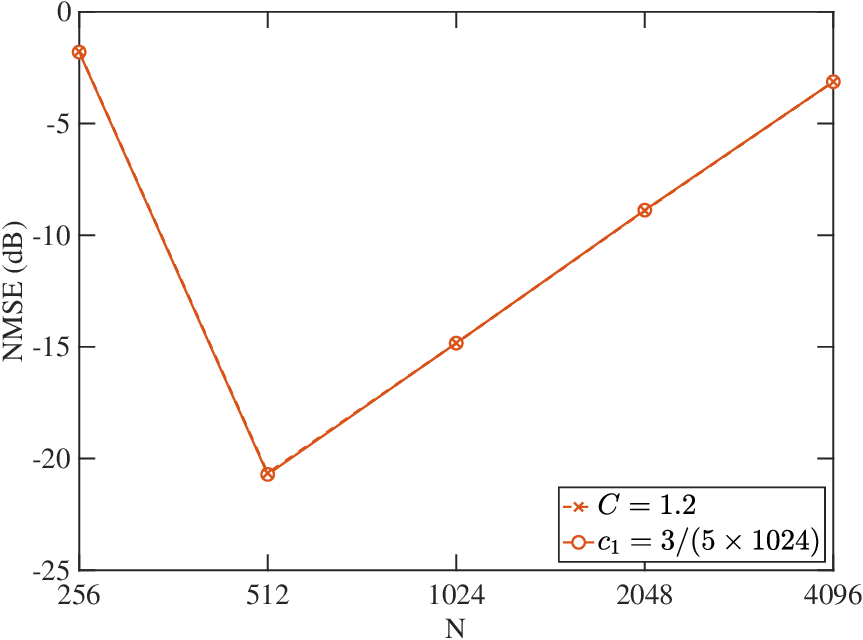}
  \caption{NMSE of the DT-AFDM I/O relation versus delay-grid size $N$ for the diagnostic single-path channel at a fixed delay, comparing fixed $C$ with fixed $c_1$.}
  \label{fig:conventional_grid}
\end{figure}

Figs.~\ref{fig:conventional_eva}--\ref{fig:conventional_grid} investigate the validity regime of the conventional DT-AFDM sequence-level relation in \eqref{eq:yf} by comparing its predictions with a waveform-level transceiver simulation. Section~\ref{sec:waveforms} showed that the ideal bandlimited continuous-time interpretation of the DT-AFDM sequence corresponds to sinc pulse shaping, so the waveform-level reference employs a sinc pulse and its matched receive filter. Meanwhile, because the conventional relation admits only integer delay indices, its prediction for an off-grid channel is formed by assigning each continuous-valued path delay to its nearest delay-grid point, while the waveform-level reference retains the original continuous-valued delays.

Fig.~\ref{fig:conventional_eva} first compares the NMSE over the standard off-grid EVA channel with that over the two diagnostic channels. For quasi-EVA, the $Q=128$ curve is limited by sinc truncation and remains around $-28$~dB, whereas $Q=2048$ substantially reduces this residual, and a Doppler-dependent increase from about $-48$~dB at zero speed to $-39$~dB at 500~km/h then appears. By contrast, the off-grid EVA result remains around $-6$~dB, compared with about $-15$~dB for the single off-grid path. The weak speed dependence of these two off-grid cases implies that accumulated fractional-delay mismatches, rather than Doppler, dominate over this range of parameter settings. The same dominance makes them insensitive to the pulse span, and repeating both off-grid cases with $Q=2048$, not plotted here, changes neither value by more than $0.2$~dB.

Fig.~\ref{fig:conventional_frac_delay} uses the diagnostic single-path channel to isolate fractional delay by sweeping the path delay over $(10+\delta)T/N$, where $-0.5\leq\delta\leq0.5$, at 500~km/h with a matched sinc pulse pair of span $Q=128$. Since the DAFT is unitary, the DAFT-domain NMSE in \eqref{eq:nmse} equals its sampled-time-domain counterpart. For independent unit-power symbols, this NMSE can be evaluated from the equivalent-tap energies. At $\nu_p=0$, let $\kappa=l-10$. The derived relation gives $h_p\sinc(\kappa-\delta)$, whereas the conventional relation assigns $h_p$ only to $\kappa=0$. Using $\sum_\kappa\sinc^2(\kappa-\delta)=1$, we obtain
\begin{align}\label{eq:frac_delay_nmse}
  \mathrm{NMSE}
   & = \left(1-\sinc(\delta)\right)^2 + \sum_{\kappa\neq 0}\sinc^2(\kappa-\delta) \nonumber \\
   & = 2\left(1-\sinc(\delta)\right),
\end{align}
where the two terms represent the mismatch at the retained tap and the energy in the remaining taps, respectively.

Derived at $\nu_p=0$, \eqref{eq:frac_delay_nmse} nevertheless closely predicts the NMSE observed at 500~km/h, with Doppler fully retained in the waveform simulation. In \eqref{eq:tv_tap_mf}, the time-dependent Doppler phase, $e^{j2\pi\nu_p k'T/N}$, describes the path variation across received samples indexed by $k'$. Recall that $\bar\nu_p=\nu_pT/N$. For $\tau_p=(10+\delta)T/N$, define $\xi=(\kappa-\delta)T/N$. The ambiguity kernel for an untruncated sinc pair is then
\begin{align}\label{eq:sinc_kernel}
  \mathcal A_{a,a}(\xi,-\nu_p)
  ={}&(1-|\bar\nu_p|)
  \sinc\!\left((1-|\bar\nu_p|)(\kappa-\delta)\right) \nonumber \\
  &\times e^{-j\pi\bar\nu_p(\kappa-\delta)},
\end{align}
where the tap-dependent Doppler phase, $e^{-j\pi\bar\nu_p(\kappa-\delta)}$, varies across the relative delay taps indexed by $\kappa$ and is independent of the received-sample index $k'$. Since $|\bar\nu_p|\leq6\times10^{-4}$, this phase remains close to unity over the taps carrying significant energy and therefore has negligible influence on the NMSE. Doppler also acts on the kernel magnitudes, and there the pulse span enters. The factor $1-|\bar\nu_p|$ reduces the retained tap, while the same rescaling of the sinc argument displaces its zeros from the tap grid, so that every remaining tap acquires a contribution of order $|\bar\nu_p|$. These contributions do not decay with the tap index, so their total is governed by the number of taps the kernel spans rather than by its shape near the peak, and for $Q=128$ it remains well below the residual left by sinc truncation. The same accumulation, larger for the longer kernel, is what produces the speed dependence of the quasi-EVA curves in Fig.~\ref{fig:conventional_eva}. Consequently, the waveform-simulation results at 500~km/h closely follow \eqref{eq:frac_delay_nmse}, reaching about $-7$~dB at $|\delta|=0.25$ and $-1.4$~dB at $|\delta|=0.5$. Only near $\delta=0$, where the fractional-delay mismatch vanishes, does the residual become visible, and it is then set by sinc truncation rather than by Doppler.

\begin{figure}
  \centering
  \includegraphics[width=8.5cm]{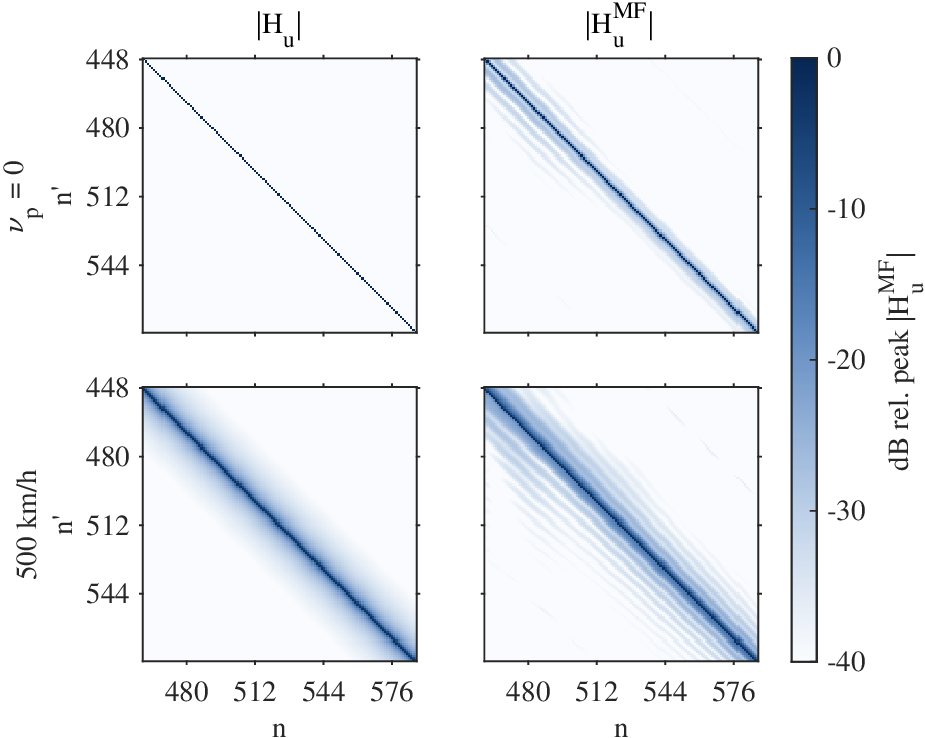}
  \caption{Magnitudes of $\mathbf H_u$ and $\mathbf H_u^{\rm MF}$ for the diagnostic single-path channel, shown in dB relative to the peak of $|\mathbf H_u^{\rm MF}|$. Rows $448$--$575$ and columns $460$--$587$ of the $1024\times1024$ matrices are displayed.}
  \label{fig:hu_heatmap}
\end{figure}

Although the validity-regime results in Figs.~\ref{fig:conventional_eva}--\ref{fig:hu_heatmap} use the sinc pulse pair, similar small-normalized-Doppler reasoning applies to an SRRC pair. Let $T_{a,\mathrm{eff}}$ denote the effective overlap duration over which $a(t)a^*(t-\xi)$ is appreciable. This duration is typically a few times the pulse main lobe width, which scales with the sampling interval $T/N$. Under the small-Doppler condition $2\pi|\nu_p|T_{a,\mathrm{eff}}\ll1$, $\mathcal A_{a,a}(\xi,-\nu_p)\approx\mathcal A_{a,a}(\xi,0)=R_a(\xi)$, where $R_a(\xi)$ is the pulse autocorrelation. This delay-only approximation, adopted in \cite{shen_fafdm_arxiv_2026}, retains the time-dependent Doppler phase while replacing the two-dimensional ambiguity kernel by the pulse autocorrelation. When the Doppler-induced phase variation over the effective overlap duration cannot be neglected, the full two-dimensional ambiguity kernel must be retained, consistent with the Doppler-dependent spreading described in Section~\ref{sec:io_relation}. That condition governs the shape of the kernel rather than the residual accumulated across taps, which for a practical SRRC span extends over far fewer taps than for the sinc spans used here.

Fig.~\ref{fig:conventional_grid} uses the diagnostic single-path channel to examine whether refining the delay grid removes the fractional-delay-induced mismatch. Here, $T$ and the delay $\tau=9.9T/1024$ are fixed, while $N$ varies from 256 to 4096. The speed is 500~km/h, and the sinc span is $Q=2048$. The first curve sets $c_1=\frac{3}{5N}$, keeping $C=2N|c_1|=1.2$, whereas the second fixes $c_1=\frac{3}{5\times1024}$, so that $C$ varies from $0.3$ to $4.8$. Because the delay is fixed, both cases have the same signed fractional displacement from the nearest grid point, defined as $\delta_N=\tau N/T-\left\lfloor\tau N/T+1/2\right\rfloor$. This centered modulo-one quantity takes the values $0.475$, $-0.05$, $-0.1$, $-0.2$, and $-0.4$ as $N$ increases. The two NMSE curves are nearly superposed, showing that the mismatch is governed by the fractional displacement $\delta_N$ rather than by $C$. Neither curve decreases monotonically with $N$, because the NMSE follows $|\delta_N|$, which does not shrink as the grid is refined. Delay-grid refinement therefore does not remove the fractional-delay mismatch.

\subsubsection{Structure of the Channel Matrices}
Fig.~\ref{fig:hu_heatmap} reveals the matrix structure of the mismatch by comparing the conventional DT-AFDM channel matrix $\mathbf H_u$ in \eqref{eq:yf} with the PS-AFDM channel matrix $\mathbf H_u^{\rm MF}$ in \eqref{eq:daft_io} for the diagnostic single-path channel. The former is calculated directly from the conventional sequence-level relation, whereas the latter accounts for matched sinc pulse shaping and receive filtering with $Q=128$. The two matrices exhibit markedly different coupling patterns. At $\nu_p=0$, $\mathbf H_u$ concentrates all its energy on the circulant offset $n'-n=-2Nc_1l_p=-12$, as implied by its delta kernel, whereas $\mathbf H_u^{\rm MF}$ spreads the path contribution across a band governed by the shifted-sinc ambiguity kernel. Fractional Doppler at 500~km/h further spreads both matrices, yet the relative Frobenius discrepancy between the full matrices, which the displayed crop does not show, is about $-15$~dB at both zero and 500~km/h. This value agrees with \eqref{eq:frac_delay_nmse} at $|\delta|=0.1$, linking the observed matrix discrepancy to the analytical fractional-delay mismatch.

Taken together, these results show that the conventional DT-AFDM sequence-level relation is exact when every physical path has an on-grid delay and zero Doppler and the sampled ambiguity kernel reduces to a delta function, whereas finite-span pulse truncation, fractional delays, or non-negligible Doppler can introduce a substantial mismatch. Because physical propagation delays are continuous-valued, exact alignment with the sampling grid generally does not occur. The relation may remain useful as an approximation when the fractional displacement and $\bar\nu_p$ are both small, but the mismatch increases as either effect becomes appreciable. Delay-grid refinement does not necessarily eliminate this mismatch.

\subsection{Waveform-Generation Complexity Comparison}
We compare the waveform-generation complexity of PS-AFDM with that of ODDM.
Note that the PS-AFDM signal occupies a time-bandwidth product of approximately $T \times \frac{N}{T} = N$. This provides $N$ degrees of freedom (DoF) to carry $N$ digital symbols, which is consistent with classical communication theory \cite{tse_fwc_2005,madhow_fdc_2008,jacobs_pce_1965}. This also explains why the DT-AFDM sequence consists of sub-Nyquist samples: the AFT-MC/CT-AFDM waveform occupies excessive bandwidth for transmitting only $N$ symbols, leading to spectral efficiency well below the available DoF.
For a fair comparison, the number of subcarriers $N_{\rm OD}$ and the number of symbols $M_{\rm OD}$ in ODDM are configured as $M_{\rm OD}N_{\rm OD}=N$, so that the two systems transmit $N$ digital symbols with the same subcarrier spacing $\frac{1}{T}$, while occupying nearly the same TF region.

The complexity is evaluated in terms of the transform operations and filtering procedures required for waveform generation only.
It is known that the ODDM waveform can also be \emph{approximately} realized by generating $N$ time-domain samples and subsequently passing them through a root-Nyquist pulse-shaping filter \cite{lin_oddm_twc_2022,lin_primer_spawc_2025,lin_mcevo_tcom_2026}.
Therefore, the comparison focuses on the generation of the $N$ time-domain samples.

As shown above, PS-AFDM is generated from an $N$-subcarrier DT-AFDM sequence with IFFT complexity $\mathcal{O}(N\log N)$. By comparison, ODDM is an $N_{\rm OD}$-subcarrier system requiring $M_{\rm OD}$ IFFTs of size $N_{\rm OD}$, resulting in a complexity of
$
  M_{\rm OD}\mathcal{O}(N_{\rm OD}\log N_{\rm OD})
  =  \mathcal{O}(N\log N_{\rm OD}),
$
which is substantially lower than $\mathcal{O}(N\log N)$, since typically $N_{\rm OD} \ll N$.
Furthermore, it is known that a CP is not necessary for ODDM signals \cite{pan_cpfreeoddm_twc_2026}, as ODDM is essentially a pseudo-impulse-based transmission scheme for DD channels \cite{lin_primer_spawc_2025}.
Consequently, when transmitting the same number of digital symbols with the same subcarrier spacing (frequency resolution), the waveform-generation complexity of ODDM is notably lower than that of PS-AFDM.

Note that a receiver-side comparison cannot yet be made. Since $\mathbf{H}_u^{\rm MF}$ lacks the pure path-wise DD structure assumed by the conventional model, the corresponding channel estimators and subsequent equalizers remain to be developed.

\section{Conclusion}
\label{sec:conclusion}

This paper establishes a unified continuous-time framework for chirp-domain waveforms, connecting their WH representation, spectral properties, sampled implementations, and practical I/O behavior. It interprets AFT-MC/CT-AFDM as PS-OFDM waveforms, explains the aliasing induced by sub-Nyquist sampling and the resulting conditional orthogonality of the I-AFDM waveform, and shows that sample-wise root-Nyquist pulse shaping yields the transmittable PS-AFDM waveform.
The derived PS-AFDM I/O relation captures the complete transmit-channel-receive chain through a cross-ambiguity kernel and shows that the effective channel is generally not a superposition of pure path-wise DD components. Analysis and waveform simulations characterize the restricted validity regime of the conventional DT-AFDM relation and demonstrate substantial mismatch for practical continuous-valued delays.
The relation also indicates what the corresponding receiver design must address. Since a physical path contributes to several equivalent taps rather than one, pilot patterns have to be dimensioned against a spread support, and estimation has to recover a channel parameterized by the ambiguity kernel rather than independent on-grid coefficients. The kernel is fixed by the pulse pair and is therefore known at the receiver, so the unknowns remain the physical path parameters, a smaller set than the taps they induce. Whether this structure can offset the loss of path-wise DD sparsity in equalization is left for future work.

\appendices

\section{Approximate Bandwidth of \texorpdfstring{$g_{\rm AF}(t)$}{g\_AF(t)}}
\label{sec:derivation_of_bandwidth}

Evaluating the Fourier integral in \eqref{eq:root-chirp-spectrum} for $c_1>0$ by completing the square gives
\begin{align*}
  G_{\rm AF}(f)=\frac{T}{\varrho}\,e^{-j\pi f^{2}T/B_{0}}\left[\mathcal{E}(w_{1})-\mathcal{E}(w_{0})\right],
\end{align*}
where $\mathcal{E}(w)\triangleq\int_{0}^{w}e^{j\pi v^{2}/2}\,dv$ is the complex Fresnel function, $\varrho\triangleq 2N\sqrt{c_{1}}$, $B_{0}\triangleq 2c_{1}N^{2}/T$, and
\begin{align*}
  w_{1}(f)\triangleq\varrho\left(1-\frac{f}{B_{0}}\right),\quad
  w_{0}(f)\triangleq-\frac{\varrho f}{B_{0}}.
\end{align*}
The exponential factor has unit magnitude, so the spectral shape is set by $\mathcal{E}(w_{1})-\mathcal{E}(w_{0})$. Applying $\mathcal{E}(w)=\operatorname{sgn}(w)\frac{1+j}{2}+O(1/|w|)$, valid for $|w|\gg1$, this difference tends to $1+j$ when $w_{1}$ and $w_{0}$ have opposite signs and vanishes when they share a sign. Since $w_{1}-w_{0}=\varrho>0$, the former occurs exactly for $0<f<B_{0}$. The main spectral region is therefore $[0,B_{0}]$, giving the approximate bandwidth $B_{g_{\rm AF}}\approx B_{0}=2c_{1}N^{2}/T$. For $c_{1}<0$, $g_{\rm AF}(t)$ is the complex conjugate of the chirp with $|c_{1}|$, so $|G_{\rm AF}(f)|$ is mirrored about $f=0$ and the main spectral region becomes $[-2|c_{1}|N^{2}/T,\,0]$, giving the same width $B_{g_{\rm AF}}\approx 2|c_{1}|N^{2}/T$.

This is an approximation rather than a strict support. The asymptotic expansion fails near $f=0$ and $f=B_{0}$, where one argument approaches zero and a finite-width Fresnel transition region arises, and the truncation of the chirp by $\Pi_{T}(t)$ leaves decaying sidelobes outside $[0,B_{0}]$.

\section{Orthogonality Analysis of Aliased Chirps}
\label{sec:orthogonality}

Recall that the aliased chirp is given by
\begin{align*}
  \hat \phi_n(t)
  =e^{j2\pi\left(c_2 n^2 + \frac{c_1}{\Delta t^2}t^2 + \frac{n}{T}t - \frac{q}{\Delta t}t\right)},
\end{align*}
where $T = N\Delta t$, $C = 2N|c_1|$, and the interval index $q$ updates at boundaries $t_{n,q} = \frac{T}{C}q - \frac{n}{C}\Delta t$.
This boundary condition allows us to express $q$ as an explicit function of $t$, denoted $q_n(t)$, using the floor function
\begin{align*}
  q_n(t) = \left\lfloor \frac{C}{T}t + \frac{n}{N} \right\rfloor.
\end{align*}
Then, the inner product
\begin{align*}
  I_{n,n'} & = \int_{0}^{T} \hat \phi_n(t) \hat \phi_{n'}^*(t) dt \nonumber                                                      \\
           & = e^{j2\pi c_2(n^2 - n'^2)} \int_{0}^{T} e^{j2\pi \left( \frac{n''}{T}t - \frac{\Delta q(t)}{\Delta t}t \right)} dt,
\end{align*}
where $n''=n-n'$ and $\Delta q(t) = q_n(t) - q_{n'}(t)$.

We first consider integer $C$. In this case, the integration interval $[0,T]$ can be partitioned into $C$ segments of length $T/C$. Introduce a local time variable $\varepsilon\in[0,1)$ and an interval index $\hat c\in\{0,1,\dots,C-1\}$ such that
$t = (\varepsilon + \hat c)\frac{T}{C}$.

Substitute this into $\Delta q(t)$. Because $\hat c$ is an integer, we can use the floor function property $\lfloor x + \hat c \rfloor = \lfloor x \rfloor + \hat c$ to separate the additive term $\hat c$: 
\begin{align*}
  \Delta q(t) & = \left\lfloor (\varepsilon+\hat c) + \frac{n}{N} \right\rfloor - \left\lfloor (\varepsilon+\hat c) + \frac{n'}{N} \right\rfloor \nonumber                               \\
              & = \left( \left\lfloor \varepsilon + \frac{n}{N} \right\rfloor + \hat c \right) - \left( \left\lfloor \varepsilon + \frac{n'}{N} \right\rfloor + \hat c \right) \nonumber \\
              & = \left\lfloor \varepsilon + \frac{n}{N} \right\rfloor - \left\lfloor \varepsilon + \frac{n'}{N} \right\rfloor \equiv \Delta q(\varepsilon).
\end{align*}
The integer $\hat c$ completely cancels out of the expression. Because the resulting expression depends solely on the local time variable $\varepsilon$ and is independent of $\hat c$, we introduce the shorthand $\equiv \Delta q(\varepsilon)$. This explicitly denotes that $\Delta q$ is periodic across every segment $\hat c$.

Substituting $t = (\varepsilon + \hat c)\frac{T}{C}$ into the integral and rewriting as a sum over $\hat c$, we obtain
\begin{align*}
  I_{n,n'} \propto \frac{T}{C} \sum_{\hat c=0}^{C-1} \int_{0}^{1} e^{j2\pi \left( \frac{n''}{T} - \frac{\Delta q(\varepsilon)}{\Delta t} \right) (\varepsilon+\hat c)\frac{T}{C}} d\varepsilon.
\end{align*}
Using $T = N\Delta t$, the phase simplifies to $\frac{n'' - N\Delta q(\varepsilon)}{C}(\varepsilon+\hat c)$. We define the integer function $\eta (\varepsilon) = n'' - N\Delta q(\varepsilon)$, allowing us to separate the variables:
\begin{align}\label{eq:geometric_sum}
  I_{n,n'} \propto \frac{T}{C} \int_{0}^{1} e^{j2\pi \frac{\eta(\varepsilon)}{C} \varepsilon} \underbrace{\left[ \sum_{\hat c=0}^{C-1} e^{j2\pi \frac{\eta(\varepsilon)}{C} \hat c} \right]}_{\text{Geometric Sum } S(\varepsilon)} d\varepsilon,
\end{align}
where
$\eta(\varepsilon) = (n-n') - N \left( \left\lfloor \varepsilon + \frac{n}{N} \right\rfloor - \left\lfloor \varepsilon + \frac{n'}{N} \right\rfloor \right)$ is an integer with $|\eta(\varepsilon)| \le N-1$.

One can see that the orthogonality depends on the geometric sum $S(\varepsilon)$ in \eqref{eq:geometric_sum}. The sum evaluates to $C$ if $\eta(\varepsilon)$ is a multiple of $C$, and $0$ otherwise.

\begin{figure*}
  \centering
  \includegraphics[width=5.7cm]{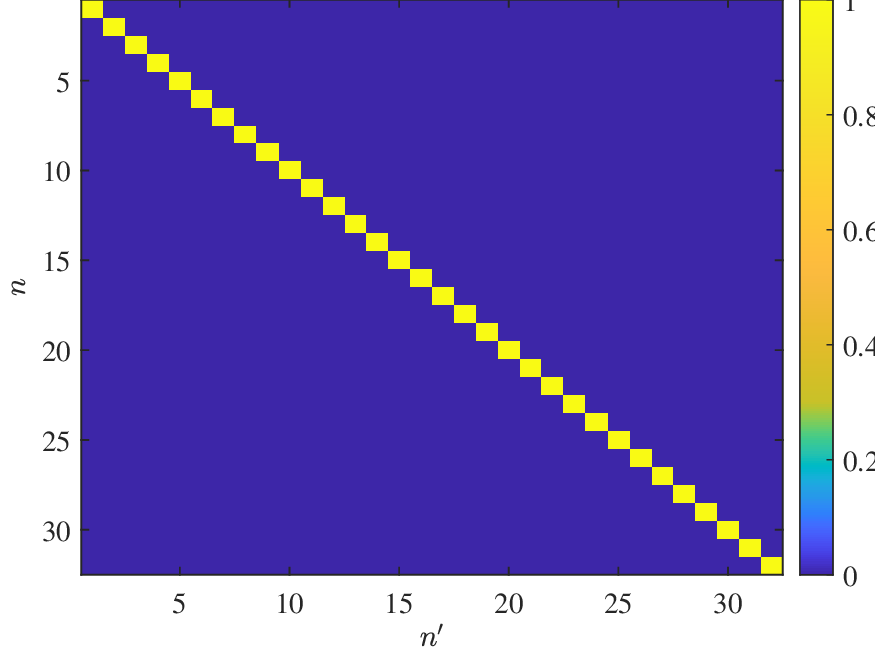}\hfill
  \includegraphics[width=5.7cm]{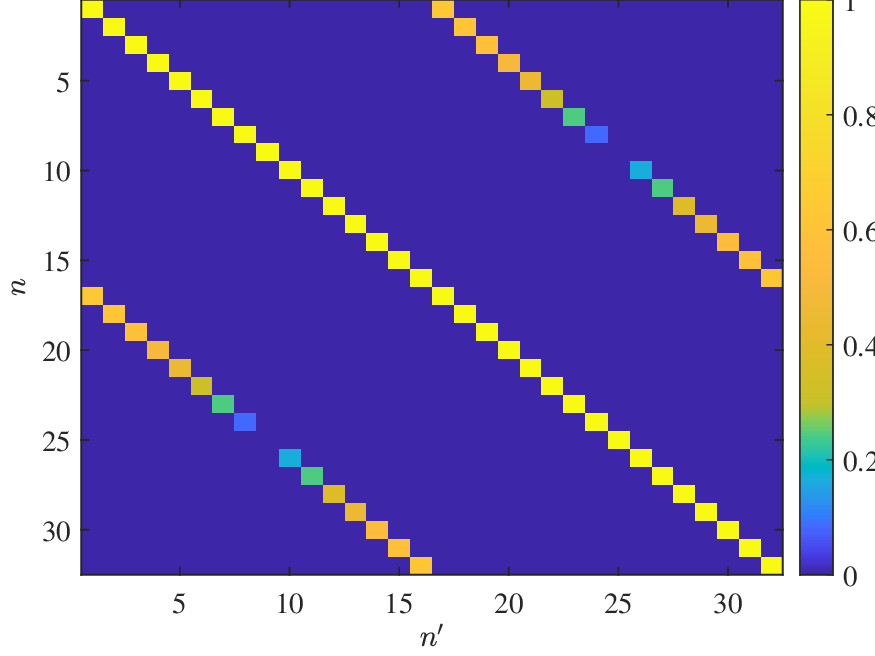}\hfill
  \includegraphics[width=5.7cm]{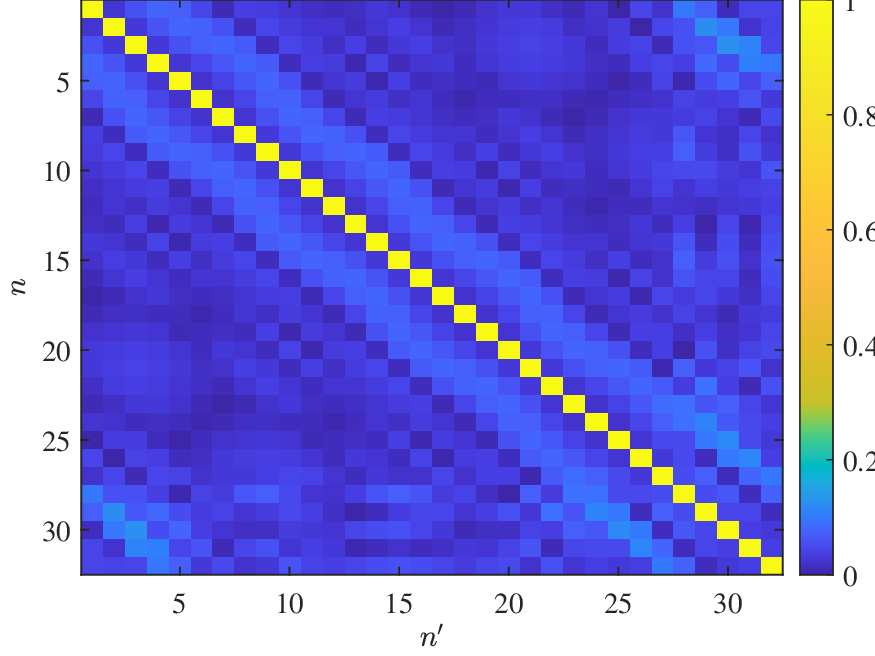}
  \caption{$|I_{n,n'}|$ for $N=32$. Left: $C=48$ or $C=32$, Case 1. Middle: $C=16$, Case 2. Right: $C=1.2$, Case 3.}
  \label{fig:N32_orth}
\end{figure*}

\subsubsection*{Case 1: $C\in\mathbb Z^{+}$ and $C \ge N$}
If $C \ge N$, then $|\eta(\varepsilon)| < C$. Since $n \neq n'$, $\eta(\varepsilon)$ cannot be $0$. Therefore, $\eta(\varepsilon)$ is never a multiple of $C$. Perfect orthogonality holds.

\subsubsection*{Case 2: $C\in\mathbb Z^{+}$ and $C < N$}
Because $|\eta(\varepsilon)|$ can reach $N-1$, it is now possible for $\eta(\varepsilon) = \pm C$. When this occurs for specific pairs of chirps for which $n-n'$ is a multiple of $C$, $e^{j2\pi \frac{\eta(\varepsilon)}{C}} = 1$. Orthogonality fails and aliasing occurs between chirps separated by index $C$.

\subsubsection*{Case 3: $C\notin\mathbb Z^{+}$}
Although the piecewise I-AFDM definition in Section~\ref{sec:iafdm} implicitly assumes integer $C$, general AFDM parameter settings need not satisfy this restriction. Let $J=\lfloor C\rfloor$ and $r=C-J$, where $0<r<1$. The interval $[0,T]$ then consists of $J$ complete segments and one residual segment, giving
\begin{align}
  I_{n,n'}
  &=\frac{T}{C}e^{j2\pi c_2(n^2-n'^2)}
  \Bigg[
  \int_0^1 e^{j2\pi\eta(\varepsilon)\varepsilon/C}
  \sum_{\hat c=0}^{J-1}
  e^{j2\pi\eta(\varepsilon)\hat c/C}\,d\varepsilon \nonumber\\
  &\qquad+
  \int_0^r
  e^{j2\pi\eta(\varepsilon)(J+\varepsilon)/C}\,d\varepsilon
  \Bigg].
  \label{eq:noninteger_C_inner_product}
\end{align}
Because the geometric sum contains only $J=\lfloor C\rfloor$ terms, it does not necessarily form a complete root-of-unity sum. The residual interval also contributes an additional term. Hence, the complete root-of-unity cancellation available for integer $C$ does not generally occur, and orthogonality for noninteger $C$ is parameter dependent and must be determined from \eqref{eq:noninteger_C_inner_product}.

Let $\mathbf I$ denote the $N\times N$ matrix with entries $I_{n,n'}$. Fig.~\ref{fig:N32_orth} shows the heatmap of $|I_{n,n'}|$ for $N=32$, using representative values of $C$ for the three cases. For Case 1, where $C=48$ and $C=32$, $\mathbf I$ is diagonal, confirming mutual orthogonality of the aliased chirps. In Case 2, with $C=16<N$, two off-diagonal bands appear at $|n-n'|=16$, indicating structured interference between chirps separated by $C$. For the noninteger setting $C=1.2$, there are nonzero off-diagonal entries, in agreement with \eqref{eq:noninteger_C_inner_product}.

\section{Derivation of the Equivalent Channel Tap \texorpdfstring{$h_{k',l}^{\rm MF}$}{h\_kl\_MF}}
\label{sec:derivation_of_channel_tap}

The channel tap $h_{k',l}^{\rm MF}$ can be interpreted as the channel observed when a discrete sequence with sample interval $T_s$ is pulse-shaped by $a(t)$, transmitted through the DD channel and subsequently processed by a matched filter.

Let us consider the transmitted waveform $x(t) = \sum_k x[k] a\left(t - kT_s\right)$, and substitute it into \eqref{eq:rt_channel} to yield
\begin{align*}
  y(t)
   & = \sum_{p=1}^{P}\sum_k h_p x[k] e^{j2\pi \nu_p t} a(t-\tau_p-kT_s),
\end{align*}
where the noise term is ignored.
The output of the matched filter is given by
\begin{align*}
  y^{\rm MF}(t)
   & = \int_{-\infty}^{\infty} y(\lambda)a^*(\lambda-t)d\lambda  \\
   & = \sum_{p=1}^{P}\sum_k h_p x[k] \nonumber \\
   & \quad \times
  \int_{-\infty}^{\infty}
  e^{j2\pi \nu_p \lambda}
  a(\lambda-\tau_p-kT_s)
  a^*(\lambda-t)
  d\lambda.
\end{align*}
Applying the change of variables $\xi=\lambda-\tau_p-kT_s$ leads to
\begin{align*}
  y^{\rm MF}(t)
   & = \sum_{p=1}^{P}\sum_k h_p x[k]
  e^{j2\pi \nu_p(\tau_p+kT_s)}                          \\
   & \quad \times
  \int_{-\infty}^{\infty}
  a(\xi)a^*(\xi-(t-\tau_p-kT_s))
  e^{j2\pi \nu_p \xi}
  d\xi                                                       \\
   & =\sum_{p=1}^{P}\sum_k h_p x[k]e^{j2\pi \nu_p t}
  \mathcal A_{a,a}(\hat \tau, -\nu_p),
\end{align*}
where $\hat \tau = t-\tau_p-kT_s$.
Then, sampling at $t=k'T_s$ and defining the delay tap index $l\triangleq k'-k$, we obtain
\begin{align*}
  y^{\rm MF}[k']
   & \triangleq y^{\rm MF}(k'T_s) \nonumber \\
   & =
  \sum_l \sum_{p=1}^{P}
  h_p e^{j2\pi \nu_p k'T_s} \nonumber              \\
   & \quad \times
  \mathcal A_{a,a}(lT_s-\tau_p,-\nu_p)
  x[k'-l].
\end{align*}
Accordingly, the equivalent channel tap is given by
\begin{align*}
  h_{k',l}^{\rm MF}
  \triangleq
  \sum_{p=1}^{P} h_p
  e^{j2\pi \nu_p k'T_s}
  \mathcal A_{a,a}(lT_s-\tau_p,-\nu_p).
\end{align*}
Substituting $T_s=\frac{T}{N}$ completes the derivation.

\bibliographystyle{IEEEtran}
\bibliography{ref26,ocdm}

\end{document}